\documentstyle[aps,prl,multicol,epsf]{revtex}
\def\gs{\gtrsim}
\def\ls{\lesssim}
\def\be{\begin{equation}}
\def\en{\end{equation}}                  
\def\p{\partial} 
\newcommand{\bi}[1]{\mbox{\boldmath$#1$}}
\newcommand{\av}[1]{\langle{#1}\rangle}
\newcommand{\avs}[1]{{\langle{#1}\rangle}_{\rm s}}

\def\bea{\begin{equation}\begin{array}{rcl}}
\def\ena{\end{array}\end{equation}}
\newcommand{\ppp}[3]{{\bigg(}\frac{\partial {#1}}{\partial {#2}}{\bigg )}_{#3}}

\begin{document}
\draft
\bibliographystyle{prsty}
\title{ Convective Heat Transport  in  Compressible Fluids}
\author{ Akira Furukawa and Akira  Onuki}
\address{Department of Physics, Kyoto University, Kyoto 606-8502}
\date{\today}
\maketitle

\begin{abstract} 
We present  hydrodynamic equations 
of compressible fluids  in gravity as  a 
generalization of those 
in the Boussinesq approximation used for nearly 
incompressible fluids. 
They account for adiabatic 
processes taking  place throughout 
the cell (the piston effect) and those 
taking place within plumes 
(the adiabatic temperature gradient effect).
Performing two-dimensional numerical analysis, 
we reveal some unique features 
of plume generation and 
convection in  transient and steady states 
of compressible fluids. 
As  the critical point is approached, 
 overall temperature changes induced by 
plume arrivals at the 
boundary walls are amplified, giving rise 
 to overshoot behavior 
in  transient states and  significant 
noises of the temperature 
in  steady states. 
 The velocity  field is  suggested to 
 assume a logarithmic 
 profile within boundary layers. Random reversal of macroscopic 
 shear flow is 
  examined in a cell with unit aspect ratio. 
We also present a  simple scaling theory for  
moderate Rayleigh numbers.

\end{abstract}
\pacs{PACS numbers: 44.25.+f, 47.27.Te, 64.70.Fx}

\begin{multicols}{2}

\date{ }
\pagestyle{empty}

\section{Introduction}

Recently much attention has been paid 
to organized fluid motion 
in turbulent convection in the 
Rayleigh-B$\acute{\rm e}$nard geometry 
\cite{Chan,SiggiaR,Siggia,Kadanoff,Ca,Gross,Gro}. 
Although the conventional hydrodynamic equations 
are constructed for (nearly) incompressible fluids 
\cite{one}, we may mention 
a number of  convection experiments 
in compressible one-component  fluids in the 
supercritical region 
\cite{Sano,Wu,Ash,Thesis,Cha,Ahlers,Ahlers-Xu,MeyerPRL,MeyerPRE,N0,N1},
 together with those in non-critical fluids such as 
water or Hg \cite{Gollub,Cioni,Tong}. 
In these studies  the Nusselt number $Nu$ representing 
the efficiency of convective heat transport 
has been measured at large values of  
the Rayleigh  number $Ra$ defined by 
\be 
Ra= \alpha_p\rho g L^3 \Delta T/\eta D.  
\en 
Here $g$ is the gravity constant, 
$\Delta T=T_{\rm bot}-T_{\rm top}$ is the 
difference between the bottom and top temperatures,  and  
$L$ is the cell height. 
As the critical point 
is approached in one-component  fluids,  
the thermal expansion coefficient 
$\alpha_p = -(\p \rho /\p T)_p/\rho$ 
grows strongly as 
$\xi^{\gamma/\nu}$ 
(in the same manner as  the 
isothermal expansion coefficient 
$K_T= (\p \rho/\p p)_T/\rho$ and 
the isobaric specific heat $C_p$), 
the thermal diffusivity $D$ 
decreases as $\xi^{-1}$,  and   
the shear viscosity $\eta$ 
is nearly a constant.  
Here $\xi$ 
is the thermal correlation length 
growing as $(T/T_c-1)^{-\nu}$ 
on the critical isochore with 
$\gamma\cong 1.24$ and $\nu \cong 0.625$.  
Hence, in the critical region,   
$Ra$  can be  extremely large; for example, 
$Ra \sim 10^{13}$  even for not 
very long $L$($\ls 10$ cm). 
The   Prandtl number 
$Pr= \eta/\rho D$ was 
 in the range of 1-100.

High compressibility 
of  supercritical fluids  gives 
rise to some unique features  
not encountered in  incompressible fluids.
(i) First, 
the transient behavior  
after application of a heat flux from the bottom 
 is strongly influenced by  the 
so-called piston effect 
\cite{Onukibook,Onuki_Ferrell,Straub,Boukari,Zappoli,Guenoun,Zhong,Sakir}, 
as revealed by 
recent high-precision experiments on $^3$He 
\cite{MeyerPRE} 
and reproduced by subsequent simulation \cite{Chiwata}. (ii) 
Second,  as $\alpha_p$ grows,  
the usual  mechanism of convection onset 
$Ra>Ra_{\rm c}(\cong 1708)$ 
is replaced by  that of 
 the  Schwarzschild criterion \cite{Gitterman,Carles}. 
That is,  for large compressible  fluid 
columns (even far from the critical point), 
convection sets in   when 
 thermal plumes continue to rise upward 
adiabatically.   
This occurs when the applied 
temperature gradient $|dT/dz|$ 
is larger than the adiabatic gradient \cite{Landau_h}, 
\be 
a_{\rm g} = (\p T/\p p)_s \rho g, 
\label{eq:1.2}
\en   
which is equal to  $0.034$ mK$/$cm for $^3$He 
and $0.27$ mK$/$cm for CO$_2$. 
This is the condition that the  
entropy   $s=s(T,p)$  per unit mass  decreases with height 
as ${d s}/{dz}= ({C_p}/T) 
 [ {dT}/{dz}+ a_{\rm g}  ]<0$,   
under which the entropy of  fluid elements adiabatically 
convected upward is larger than that 
of the ambient fluid. More precisely,  
Gitterman and  Steinberg  \cite{Gitterman}
found that the convection onset for  
compressible fluids   is 
given by  $Ra^{\rm corr}>Ra_{\rm c}$, 
where $Ra^{\rm corr}$ is a  corrected Rayleigh number 
defined by       
\bea 
Ra^{\rm corr} &=&  
 (\alpha_p\rho g L^3 /\eta D )(\Delta T- a_{\rm g} L) \nonumber \\
&=& Ra (1-a_{\rm g}L/\Delta T) . 
\label{eq:1:3}
\ena 
This is a natural consequence because the effective temperature 
gradient seen by the raising plumes is given by $\Delta T/L-a_{\rm g}$.
At the convection onset  we thus have 
\be 
(\Delta T)_{\rm on}  =  
a_{\rm g} L  + Ra_{\rm c}{D\eta_0}/({g\rho \alpha_pL^3}) . 
\label{eq:1.4}
\en 
where the second term behaves as 
 $(T/T_c-1)^{\gamma+\nu}L^{-3}$ 
and  can be much smaller than 
the first term even for small $L (\sim 1$mm \cite{MeyerPRL}) 
as $T \rightarrow T_c$. 
The relation (1.4) has been 
confirmed in SF$_6$  \cite{Ash},  
and  in $^3$He \cite{MeyerPRL}.  
(iii) Third, in steady convective 
 states,  experimental curves of  
$Ra (Nu-1)$ vs 
${Ra}^{\rm corr}$ 
were  collapsed onto  a single universal curve 
for various densities above 
 $T_{\rm c}$  \cite{Thesis} 
and for   various average reduced temperatures
 on the critical isochore  
\cite{MeyerPRL}.  These 
empirical results are  highly nontrivial, 
because $Nu$ can in principle depend on 
$Ra$, $Pr$, and $a_{\rm g}L/\Delta T$, while   
$Nu$ is a function of $Ra$ and $Pr$ 
for   incompressible fluids.

For various fluids under relatively large 
$\Delta T \gg a_{\rm g} L$ (where $Ra^{\rm corr} 
\cong Ra$), data of $Nu$ have 
been fitted to a simple scaling law,  
\be 
Nu \sim Ra^a .
\label{eq:1.5}
\en 
The exponent  $a$ has been in a range 
from 0.28 to 0.31 and, in particular, a theoretical 
value $2/7$  \cite{SiggiaR,Kadanoff} 
was generally consistent with data for $Ra \ls 10^{12}$   
\cite{Sano,Wu,Ash,Thesis,Cha,Ahlers,Ahlers-Xu,MeyerPRL,MeyerPRE,N0,Cioni}. 
 Moreover, measurements  of the 
patterns of isothermal surfaces \cite{Gollub} and 
the velocity \cite{Cioni,Tong}  
have  been informative 
on  plume motion   and a   
large-scale circulating shear flow  
in small-aspect-ratio cells \cite{N1,Cioni}. 
Several authors have also 
performed  numerical analysis of convection at large $Ra$ 
in two dimensions (2D) \cite{Werne,Hansen,Ba,Toh,Swinney} 
and in three dimensions (3D) \cite{Christ,Ver,Kerr1,Kerr2}. 
Even in 2D salient features in the experiments  
have been reproduced. In these simulations,  if the temperature 
is averaged over a long time, 
 the temperature gradient is localized 
in thin boundary layers with 
thickness $\ell_T$ related to $Nu$ by  
\be 
Nu= L/2\ell_T. 
\label{eq:1.6} 
\en 
Both in 2D and in 3D (if visualized  from side),  
 the plumes tend to be  connected 
from  bottom to  top for large $Pr$ 
because of  slow thermal diffusion, 
while they become diffuse far from the 
boundaries for small $Pr$.  
In the 3D simulations 
with  periodic or free-slip sidewalls \cite{Kerr1,Kerr2},   
local boundary shear 
flows were observed  between incoming plumes 
and  outgoing networks  of buoyant sheets 
in horizontal planes close to the boundaries.

In this paper we will  derive and examine  
hydrodynamic equations 
for compressible fluids under gravity 
in the supercritical region,  
 in which the oscillatory motion of sound 
 has been averaged out \cite{Onukibook}. 
Since  the time scale of 
convective motions 
is  much longer than 
that of the acoustic wave 
$t_{\rm ac}= L/c$ 
(typically of order $10^{-4}$s for $L \sim 1$ cm),  
such a description 
is convenient theoretically and is even  indispensable  
for numerical analysis. Our dynamic equations are  
a natural generalization of the  usual hydrodynamic equations 
\cite{one}. Our new predictions 
are unique particularly  when 
the piston effect comes into play, 
as has been demonstrated in the previous 
simulation \cite{Chiwata} relatively close to the 
convection onset. This paper will 
present 2D  simulations 
of our hydrodynamic equations 
for  much larger  $Ra^{\rm corr}$   both in transient 
and (dynamical) steady states.  Even in 
 steady states,  we will find 
 some   characteristic  features of turbulent 
 states,  which have not 
been reported in the previous 
simulations \cite{Werne,Hansen,Ba,Christ,Ver,Kerr1,Kerr2}, 
such as the logarithmic velocity profile 
of the velocity near the boundary \cite{Landau_h} 
and   random reversal of the large-scale 
circulating flow in small-aspect-ratio cells 
\cite{N1,Cioni}. We will also point out 
that individual arrivals of plumes 
at  the boundaries cause global 
temperature fluctuations  
in the cell via the piston effect. 
 The resultant noise level 
of the temperature fluctuations grows as 
the critical point is approached.

\setcounter{equation}{0}
\section{Theoretical Background}

\subsection{Hydrodynamic equations}

We consider   a supercritical fluid 
on the critical isochore  in a cell with 
the bottom plate at $z=0$ and the top plate 
at $z=L$.     The $z$ axis is taken in 
the upward direction and the total fluid volume  is fixed 
at $V$.  The temperature 
disturbance $\delta T ({\bi r},t) 
=T({\bi r},t)-T_{\rm top} $ measured from the 
temperature $T_{\rm top} $ at the top boundary   
is much smaller in magnitude than 
$T_{\rm top} -T_{\rm c}$. Hereafter  $\epsilon$ 
will be used to denote  the reduced temperature  at the 
top boundary, which satisfies 
\be 
\epsilon = T_{\rm top}/T_{\rm c}-1 \gg \Delta T/T_c.
\label{eq:2.1}
\en  
We  assume  that the gravity-induced density 
stratification is not too severe such that the 
thermodynamic derivatives 
are nearly homogeneous in the cell. 
This is satisfied when 
$|\rho/\rho_c-1| \sim  (\p \rho/\p p)_T g L \ll \epsilon^\beta$ 
with  $\beta\cong 0.33$ \cite{gra}. This condition is rewritten as    
\be 
\epsilon^{\beta+\gamma} \gg a_{\rm g}L/T_{\rm c}. 
\label{eq:2.2}
\en

In the theoretical literature on convection 
\cite{SiggiaR,Siggia,Kadanoff,Ca,Gross,Gro},  
the top and bottom temperatures $T_{\rm bot}$ 
and $T_{\rm top}$ are constant 
parameters. However, in most of  
recent convection experiments, particularly 
in cryogenic ones,  
 the heat flux at the bottom 
$Q= - \lambda (dT/dz)_{z=0}$ and 
$T_{\rm top}$  have   been  fixed.   
The  $\lambda$ is    the thermal conductivity. 
Furthermore, if  the top and bottom 
walls are   made of a metal with high 
thermal conductivity,   
the boundary temperatures  
become homogeneous 
in the lateral directions 
(unless local temperature changes 
are  too fast).  Then   
$T_{\rm bot}(t)$ and hence $\Delta T(t) $ 
are    functions  of time only. 
Metcalfe and Behringer \cite{Metcalfe}  performed 
linear stability analysis of  convection onset 
under  this {\it cryogenic}  boundary condition. 
In the nonlinear regime 
it is of great interest how 
 the boundary layer thickness 
 and the plume generation depend on  
the boundary condition.

In  equilibrium  the pressure 
 gradient is given by  $- \rho {g}\cong - \rho_{\rm c} {g}$.  
  In nonequilibrium  we  set 
 \be 
  p({\bi r},t) = p_{0}  - \rho_{\rm c} {g}z +p_1(t) + 
  p_{\rm inh}({\bi r},t) ,
\label{eq:2.3}
 \en 
 where $p_0$ is a constant, 
 $p_1(t)$ and $p_{\rm inh}$ are 
 the  homogeneous and  inhomogeneous 
 parts induced by $\delta T$, respectively. 
That is, we  assume  $\av{p_{\rm inh}}=0$, where 
  $\av{\cdots}\equiv \int d{\bi r} (\cdots)/V$ 
represents the space average.  Then $p_1$ is related to 
the space average of $\delta T$ by  
\be 
p_1(t) = (\p{p}/\p{T})_\rho \av{\delta T}(t),   
\label{eq:2.4}
\en 
which follows from  the thermodynamic relation  
$dp= (\p p/\p T)_\rho dT + 
(\p p/\p \rho)_T d\rho
$ and the condition  that 
 the space average of the density deviation vanishes 
($\av{\delta\rho} =0)$.
It is important that 
  the combination 
$p({\bi r},t)+ \rho_{\rm c} {g}z$ 
is nearly homogeneous or 
$|p_1(t)| \gg |p_{\rm inh}({\bi r},t)|$
for fluid motions  much slower than  
 the acoustic time $t_{\rm ac}= 
L/c$ ($c \sim 10^4$ cm$/$s 
 being the sound velocity) \cite{Onuki_Ferrell,homo}

Now we derive the equation for  $\delta T$ from   
the heat conduction equation 
\be 
\rho T \bigg (\frac{\p}{\p t} +{\bi v}\cdot\nabla  \bigg ) s   
=  \lambda \nabla^2 \delta T,  
\label{eq:2.5}
\en 
where $s({\bi r},t)$ is  the entropy  
 per unit mass. Here  $\delta s$    consists of the equilibrium 
part $s_{\rm eq} (z)$ with 
\be 
\frac{d}{dz} s_{\rm eq}(z)= -\ppp{s}{p}{T} \rho g= 
T^{-1}C_p a_{\rm g}
\label{eq:2.6}
\en 
and the nonequilibrium deviation,
\be
\delta s({\bi r},t) =  T^{-1}{C_p} \bigg [ 
 \delta T ({\bi r},t) - \ppp{T}{p}{s} p_1(t)  \bigg ].  
\label{eq:2.7}
\en 
With the aid of the thermodynamic 
identity $(\p T/\p p)_s=(\p T/\p p)_\rho (1- 1/\gamma_s)$, 
we rewrite     
(2.5) to obtain the desired equation for $\delta T$,   
\be 
\bigg ( \frac{\p}{\p t}+{\bi v}\cdot\nabla -D \nabla^2 \bigg ) \delta T
= -  a_{\rm g}  v_z+\alpha_s \frac{d}{d t} \av{\delta T},  
\label{eq:2.8}
\en 
where  $D=\lambda/C_p$ is    the thermal diffusivity and 
\be 
\alpha_s= 1-\gamma_s^{-1}.
\label{eq:2.9}
\en 
The specific heat ratio  $\gamma_s$ behaves as 
\be 
\gamma_s= C_p/C_V \sim \epsilon^{-\gamma+\alpha}\gg 1  , 
\label{eq:2.10}
\en 
where  
$C_p \sim \epsilon^{-\gamma}$ and 
 $C_V \sim \epsilon^{-\alpha}$ 
are  the  specific heats (per unit volume) 
at constant $p$ and $V$, 
respectively,  with $\alpha\cong 0.1$. 
The first term on the right hand side of (2.8)   arises  
from $ds_{\rm eq} /dz$.  
Inside  plumes   the temperature is adiabatically 
cooled  if they  go upward ($v_z>0$),  
or adiabatically warmed if they  go  downward $(v_z<0)$. 
In this way this term suppresses upward 
motion of warmer plumes from the bottom 
and downward motion of cooler  plumes from the top, 
resulting in the Schwarzschild criterion of convection onset 
(the adiabatic temperature gradient effect). 
On the other hand, the second    term   
arises from $p_1(t)$, leading    to  
the  piston effect \cite{Onuki_Ferrell}. 
It is worth noting  that 
the space integral of (2.8) in the cell becomes 
\be  
VC_V \frac{d}{dt}\av{\delta T} = \lambda 
\int da {\bi n}\cdot\nabla \delta T,  
\label{eq:2.11}
\en 
where use has been made of $\av{\bi v}={\bi 0}$. 
The right hand side represents  the rate of  heat supply from 
the  boundary surface, where $da$ is  surface element and 
$\bi n$ is  the outward 
surface normal.  Its  time-integration 
is the total heat supply expressed as $V\rho \av{\delta s}$, resulting in  
 \be  
C_V\av{\delta T}(t)= \rho \av{\delta s}(t) ,
\label{eq:2.12}
\en 
which also follows  (2.4) and the space average of (2.7). 
The appearance of $C_V$ in on the left hand side of (2.12) 
 is a natural consequence  under the fixed volume condition.   
Notice that  (2.7) can also be written as  
\be 
\delta T({\bi r},t) = 
 \frac{\rho T}{C_p}  \delta s ({\bi r},t) 
 + {\rho T}\bigg [ \frac{1}{C_V}-  \frac{1}{C_p} \bigg ]  \av{\delta s}(t). 
\label{eq:2.13}
\en 
This relation holds even in gravity 
if $\delta s$ is the deviation of $s-s_{\rm eq}(z)$ 
as in (2.7). In addition, the density deviation 
$\delta \rho= \rho -\av{\rho}$ is written in our 
approximation as 
\be 
\delta \rho= \rho K_T g (z-L/2) -
 \rho \alpha_p (\delta T -\av{\delta T}), 
\label{eq:2.14}
\en 
where $K_T= (\p \rho/\p p)_T/\rho$ 
and  we have set $\av{\delta\rho}=0$.

Since $C_p \gg C_V$ near the critical point,    
the   homogeneous part of $\delta T$ 
(second term) in (2.13) can easily 
dominate over the inhomogeneous  part (first term) 
even when  $\delta s$ 
is localized near a heated wall. 
 Indeed, if a thermal disturbance 
is produced within a thermal boundary layer 
with thickness  $\ell$ near the boundary, 
the ratio of the homogeneous part ($\propto \av{\delta s}$) 
to the localized inhomogeneous  part  
($\propto {\delta s}$)  in (2.13) 
is of order $(\gamma_s-1)\ell /L$ where $L$ 
is the characteristic system  length.  
Temperature homogenization 
is achieved when  $(\gamma_s-1)\ell \gg L$. 
By setting $\ell = (Dt_1)^{1/2}$ we obtain the 
time constant of this  thermal equilibration (the piston time) 
in the form,   
\be 
t_1= L^2/D(\gamma_s-1)^2.
\label{eq:2.15}
\en

Next we consider the momentum equation 
for the velocity field ${\bi v}({\bi r},t)$.  
On  long time scales, sound waves 
decay to zero  and 
the incompressibility condition  
 \be 
 \nabla\cdot{\bi v}=0
\label{eq:2.16}
 \en 
 becomes nearly satisfied  ($\gg t_{\rm homo}$)  \cite{homo}.  
 Then the dissipation  of $\bi v$ 
is   produced by the shear 
viscosity $\eta $ and  the usual 
 Navier-Stokes equation in the Boussinesq approximation 
 may be set up in the form \cite{Chan}, 
\be 
\bigg ( \frac{\p}{\p t}+{\bi v}\cdot\nabla \bigg ) {\bi v}
=- \nabla \frac{p_{\rm inh}}{\rho} +  
\alpha_p   g \delta T {\bi e}_z 
+\frac{\eta}{\rho} \nabla^2{\bi v}, 
\label{eq:2.17}
\en 
where the inhomogeneous part 
$p_{\rm inh}$ ensures 
(2.16), ${\bi e}_z$ is the unit vector along  the $z$ axis, 
and $\rho$ is the average density. 
The two equations (2.8) and (2.17)  are our 
fundamental dynamic equations closed under (2.16).  
In the conventional theory \cite{Chan,one}, 
(2.17) has been used,  but the right hand side 
of (2.8) vanishes.

As another  characteristic feature 
near the critical point,   the Prandtl number 
behaves as 
\be 
Pr=  \eta /\rho D \sim \epsilon^{-\nu}.  
\label{eq:2.18}
\en 
For example, $Pr=350$ at $T/T_{\rm c}-1=10^{-3}$ in $^3$He. 
This means that  the time scale of the thermal diffusion 
is much slower than that of the  velocity in the critical region. 
Based on   this fact, the simulation in Ref. \cite{Chiwata} 
was performed using  
the Stokes approximation  
in which the left hand side of (2.17) is set equal to zero. 
Good agreement with the experiments \cite{MeyerPRE} 
was then obtained for $Ra^{\rm corr}/Ra_{\rm c} -1 \ls 5$ 
at $\epsilon =0.05$.

For $Pr \gg 1$, let us  estimate 
the upper bound of $Ra^{\rm corr}$ 
below which  the Reynolds number $Re$ is 
smaller than 1 or the Stokes approximation is 
allowable. The characteristic 
temperature variation  $(\delta T)_{\perp}$ 
changing  perpendicularly  to the $z$ axis 
and the characteristic velocity 
field $v_{\rm pl}$ are related by 
\be  
v_{\rm pl}  \sim (\alpha_p \rho_{\rm c}  g /\eta k^2) (\delta T)_{\perp},  
\label{eq:2.19}
\en 
where  $k \sim 2\pi/L$ for roll patterns.  
If $Ra^{\rm corr}/Ra_{\rm c}$ is considerably 
(but not much) larger than 1,  $(\delta T)_{\perp}/\Delta T$ 
is of order 1 (but  somewhat smaller than 1).  
Then we obtain 
\be 
v_{\rm pl}  \sim  (Ra^{\rm corr}/ Ra_{\rm c})  D/L . 
\label{eq:2.20}
\en 
Thus the small Reynolds number regime 
is written as  
\be 
Ra^{\rm corr}/ Ra_{\rm c}  \ls Pr ,  
\label{eq:2.21}
\en 
where use has been made of $Re \sim v_{\rm pl} L\rho/\eta$. 
For $Pr \gg 1$ there is a sizable range of $Ra^{\rm corr}$ 
in which the Stokes approximation is justified. 
In passing, for $0<Ra^{\rm corr}/ Ra_{\rm c}-1 \ll 1$, 
the theory of the amplitude equation \cite{Ampli}  
predicts 
\be 
 v_{\rm pl} L/D   \sim (\delta T)_{\perp}/\Delta T \sim 
(Ra^{\rm corr}/ Ra_{\rm c}-1)^{1/2}, 
\label{eq:2.22}
\en 
from which 
we have $Nu-1 \sim 
Ra^{\rm corr}/ Ra_{\rm c}-1$ 
because 
the convective heat current 
is of order $C_p (\delta T)_{\perp} v_{\rm pl}$. 
In the next section 
we will estimate $v_{\rm pl}$ 
for much larger $Ra$.

Analogously to (2.19), the inhomogeneous 
 pressure deviation $p_{\rm inh}$ is estimated as 
$ p_{\rm inh} \sim  (\alpha_p \rho_{\rm c}  g /k) (\delta T)_{\perp}$.
If we assume $p_1(t) \sim 
(\p p/\p T)_{s} \Delta T$ from (2.4) and  
$\Delta T \sim (\delta T)_{\perp}$ as in (2.20), 
we find that $p_{\rm inh}/p_1(t)$ is of order 
$\epsilon^{-\gamma} a_{\rm g}/T_c k$ and is much smaller than 1 
from (2.2).  This estimation justifies 
the   assumption of the homogeneity of 
$\delta p({\bi r},t)+\rho_cgz$ made below (2.4).

\subsection{Free energy and heat production rate}

In the presence of small deviations of 
the temperature and  the density, $\delta T$ 
and $\delta \rho$,   around 
an reference equilibrium state, 
we have an increase of 
the free energy functional $\delta F$. Up to the bilinear order 
of the deviations,  
it is  of the form \cite{Onukibook,Landau_s}, 
\be 
\delta F =\int d{\bi r}\bigg [ \frac{C_{V}}{2T}(\delta T)^2+ 
\frac{1}{2\rho^2K_T} (\delta \rho)^2 + gz \delta \rho  \bigg ],
\label{eq:2.23}
\en  
where the  third term  is the   potential energy in gravity. 
All the deviations are assumed to change 
slowly in space compared with the thermal 
correlation length $\xi$. If we express 
$\delta\rho$ in terms of $\delta T$ as in  (2.14),  
we obtain 
\be 
\delta F =  \frac{1}{2T} 
\int d{\bi r}\bigg [ C_{p}(\delta T-\av{\delta T})^2+ 
C_{V}\av{\delta T}^2  \bigg ], 
\label{eq:2.24}
\en  
where the constant term is  omitted. 
We notice that $\delta F$ decreases dramatically 
for $\gamma_s\gg 1$ in the process  of adiabatic 
temperature homogenization. 
Furthermore, in the presence of  
velocity field, the total free energy change 
is the sum of $\delta F$ and 
the kinetic energy of the velocity field, 
\be 
F_{\rm K}  = \frac{1}{2} \int d{\bi r} \rho {\bi v}^2.  
\label{eq:2.25}
\en 
Its time derivative is calculated from 
our dynamic equations 
  (2.7) and (2.17) in the form,  
\bea 
\frac{d}{dt}( \delta F+F_{\rm K}) 
&=& -\int d{\bi r}(\epsilon_{\rm th}+\epsilon_{\rm vis}) \nonumber\\
&+& {\lambda}{T}^{-1}
 \int d{a}  [ \delta T ({\bi n}\cdot{\nabla \delta T}) ],   
\label{eq:2.26}
\ena 
where $\epsilon_{\rm th}$ and $\epsilon_{\rm vis}$ 
are the thermal and viscous heat 
production rates (per unit volume) \cite{Landau_h}, 
respectively, defined by 
\be 
\epsilon_{\rm th}
={\lambda}{T}^{-1} |\nabla \delta T|^2 , 
\label{eq:2.27}
\en
\be 
\epsilon_{\rm vis}= \eta \sum_{ij} (\p v_i/\p x_j)^2.
\label{eq:2.28}
\en   
In the second term of (2.26) the surface integral 
is  over the boundary  of the cell, 
$\bi n$ being the outward unit vector. 
In terms of 
 the heat flux from the bottom $Q$,  
it is expressed as $VQ\Delta T/TL$ if the top temperature 
is fixed.

\subsection{Basic relations in steady states}

We consider steady convective states 
in the Rayleigh-B$\acute{\rm e}$nard geometry,  
in which the flow pattern is either time-independent  
not far above  the convection onset or chaotic at larger  $Ra$. 
We treat $\Delta T$ as a constant parameter. 
Under  the  condition 
of fixed  heat flux at the bottom,  
 however, $\Delta T (t)$ exhibits 
rapidly-varying  fluctuations in chaotic states. 
In this case  $\Delta T$ in the following relations 
represents the time-average  
of $\Delta T (t)$. 
The steady state averages (over space and time) 
will be denoted by $\avs{\cdots}$ to distinguish them  
from the space averages $\av{\cdots}$ used so far.

We make (2.8) and (2.17) dimensionless by 
measuring  space and time in units of 
$L$ and $L^2/D$  and setting   $\tilde{{\bi r}}= L^{-1}{\bi r}$      
and ${\tilde t}= DL^{-2}t$. 
The temperature deviation is written as 
\be 
\delta T({\bi r},t )/\Delta T = 
1- \tilde{z} + Ra^{-1} {\cal F}(\tilde{{\bi r}}, \tilde{t}) ,     
\label{eq:2.29}
\en 
where $\tilde{z} =z/L$.    
The dimensionless function  
${\cal F}$ becomes nonvanishing 
 in convective states  and obeys 
\be 
\bigg ( \frac{\p}{\p \tilde{t}} + 
{\bi V}\cdot{\tilde \nabla}-{\tilde \nabla}^2  \bigg ) {\cal F} = 
  Ra^{\rm corr}  V_z + \alpha_s \frac{d}{d\tilde{t}} \av{{\cal F}}   , 
\label{eq:2.30} 
\en 
where  ${\tilde \nabla} = L \nabla$ is 
 the space derivative in units of $L$.
Then the (average) heat flux at the bottom is written as 
$Q= (\lambda \Delta T/L) [1+ Ra^{-1} f_\lambda]$, 
where 
\be 
f_\lambda= -  \avs{( {\p \cal{F}}/{\p {\tilde z}})_{\tilde{z}=0}}.    
\label{eq:2.31}
\en 
The $f_\lambda$ is a function of 
$Ra^{\rm corr}$  and $Pr$.   The Nusselt number 
$Nu = QL/\lambda \Delta T$ is expressed as 
\be 
Nu= 1 +  {Ra}^{-1}f_\lambda. 
\label{eq:2.32}
\en 
As the boundary condition of $\cal F$ we require 
${\cal F}=
  0$ at $\tilde{z}=0$ and $1$ 
  if $T_{\rm top}$ and $T_{\rm bot}$ are fixed. 
However,   
If  $T_{\rm top}$ and $Q$ at the bottom are fixed, 
we have ${\cal F}=0$ at ${\tilde{z}=0}$ 
and $\p {\cal F}/\p {\tilde{z}}=
 Ra(Nu-1)$   at $\tilde{z}=0$. 
The dimensionless velocity ${\bi V}(\tilde{{\bi r}}, {\tilde t}) 
=(L/D){\bi v}$ obeys 
\be 
\frac{1}{Pr}
\bigg ( \frac{\p}{\p {\tilde t}}+{\bi V}\cdot\tilde{\nabla} \bigg ) {\bi V}
=- \tilde{\nabla} P_{\rm inh} + {\cal F}  {\bi e}_z +   
{\tilde \nabla}^2{\bi V} , 
\label{eq:2.33} 
\en 
where $P_{\rm inh}$ ensures 
${\tilde \nabla}\cdot{\bi V}=0$.

Here we assume  that the piston  term, the second term  
on the right hand side of (2.30), 
can be neglected in steady states. 
For $\epsilon =0.05$,  
the piston term  in steady states is less than   a few percents 
of the convection term ${\bi v}\cdot {\tilde{\nabla}}{\cal F}$ 
in (2.30) except at the boundaries. 
It thus produces no  significant effects on  steady state heat 
transport (on $Nu$), while it can be crucial 
in the initial transient stage \cite{Chiwata}.  
Then,  if the piston term in (2.30) is neglected, 
 (2.30)  and (2.33) become 
  of the same form as those of usual  incompressible fluids 
except that $ Ra^{\rm corr}$ appears in place  of 
$Ra$.    At   much smaller  $\epsilon$, however, 
this  assumption  is questionable, 
   because   the noise part of $\av{{\cal F}}$ 
grows   as $\epsilon \rightarrow 0$,  as will be discussed 
later in the next section.  
We may conclude the following 
(at least for  $\epsilon =0.05$).
 (i) It follows the Gitterman-Steinberg criterion 
$Ra^{\rm corr} > Ra_{\rm c}$ 
in convective states in the compressible case      
 \cite{Gitterman,Carles}. 
(ii) It is more nontrivial that 
the combination 
\be 
Ra(Nu-1)= f_\lambda(Ra^{\rm corr},Pr) 
\label{eq:2.34}
\en 
should be a universal function of 
$Ra^{\rm corr}$ and $Pr$ from (2.32) 
in agreement with 
the experiments \cite{Thesis,MeyerPRL}.  Notice that 
$Ra(Nu-1)= f_\lambda(Ra,Pr)$ holds 
 for incompressible fluids  
in terms of the same $f_\lambda$. 
These  experiments 
and more decisively that by  Ahlers and Xu \cite{Ahlers-Xu}
 indicate  that $f_\lambda$ should be  nearly independent of 
$Pr$ once $Pr$ considerably exceeds 1. 
In the 3D simulation by Verzicco and Camussi \cite{Ver}, 
$Nu$ became independent of $Pr$ for $Pr\gs 0.5$. 
Theoretical support 
of this behavior using  scaling  arguments 
was  presented  in Ref.\cite{Gro}.

In steady states we may also derive 
some exact relations for variances among 
$\delta T$ and $\bi v$. 
Using the dynamic equations (2.8) and (2.17) 
we calculate the averages of $\p (\delta T)^2/\p t$,  
$\p {\bi v}^2/\p t$, and $\p (z \delta T )/\p t$ to obtain  
\be 
\avs{|\nabla \delta T|^2}= 
a_{\rm th}^2+ a_{\rm th}(a_{\rm th} -a_{\rm g}  ) (Nu-1), 
\label{eq:2.35}
\en 
\be 
\sum_{ij}\avs{(\p v_i/\p x_j)^2} = Ra (D/L^{2})^2 (Nu-1). 
\label{eq:2.36}
\en 
We also obtain a cross correlation,  
\be   
\hspace{0.4cm} \avs{v_z \delta T}= a_{\rm th} D (Nu-1),  
\label{eq:2.37}
\en 
which is nothing but the average convective heat flux 
(if $\lambda C_p$ is multiplied). 
Here  $a_{\rm th} \equiv\Delta T/L 
= -\avs{d \delta T/dz}$ 
is the average temperature gradient 
and $a_{\rm g}$ is the adiabatic temperature gradient defined by (1.2).
If  we use the usual hydrodynamic equations 
for incompressible fluids, 
the right hand side of (2.35) becomes 
$a_{\rm th}^2Nu$, while (2.36) and (2.37) 
remain the same  \cite{SiggiaR}.   
In addition,  (2.35) 
indicates $a_{\rm th}> a_{\rm g}$  
in convective states in which $Nu>1$. 
This is consistent with the convection 
criterion $Ra^{\rm corr}>Ra_{\rm c}$. 
We obtain 
the averages of the two dissipation 
rates in (2.27) and (2.28) by multiplying 
$\lambda/T$ and $\eta$ 
to (2.35) and (2.36), respectively. 
Using the thermodynamic identity 
$T\alpha_p =C_p (\p T/\p p)_s$, 
we obtain 
\be
\avs{\epsilon_{\rm th}}+\avs{\epsilon_{\rm vis}}= 
T^{-1}\lambda a_{\rm th}^2 Nu, 
\label{eq:2.38}
\en
\be
(\avs{\epsilon_{\rm th}}-T^{-1} \lambda a_{\rm th}^2)/\avs{\epsilon_{\rm vis}}
=a_{\rm th}/a_{\rm g}-1, 
\label{eq:2.39}
\en 
The first relation (2.38) also follows from  
the average  of (2.26). 
The second relation (2.39) holds only 
in convective states ($Nu>1$), while the 
right hand side is replaced by $
C_p a_{\rm th}/T\alpha_p\rho g= a_{\rm th}/a_{\rm g}$
for the usual hydrodynamic equations of incompressible fluids.

\setcounter{equation}{0}
\section{Simulation results}

We   perform  numerical 
analysis of (2.8) and (2.17) 
in 2D   using   parameters 
 of $^3$He in a cell 
with  $L=1.06$ mm. 
The reduced temperature is   
$\epsilon=0.05$ (except in Fig.13), where 
$\gamma_s= 22.8, T\alpha_p= 26.9$, $\lambda=1.88\times 10^{-4}$ 
erg/(cm$^2$s K), $D=5.42\times 10^{-5}$ cm$^2/$s, and $Pr=7.4$ 
\cite{MeyerPRL,MeyerPRE,Chiwata}.  
The condition (2.2) is well satisfied. 
The piston time $t_1$ in (2.15) is given by 
$0.42$ s.  We  apply a constant heat flux  
$Q$  at the bottom $z=0$ for $t>0$  with a fixed top temperature 
$T_{\rm top}$ at $z=L$.  In steady states  we have 
$
Ra^{\rm corr} /Ra_{\rm c} 
= 0.90  [ \Delta T/ a_{\rm g}L -1 ], 
$  
where  $a_{\rm g} L = 3.57$ $\mu$K. 
Thus $(\Delta T)_{\rm on} = 7.6$ $\mu$K  and 
$Q_{\rm on} = 13.5$ nW$/$s  at the convection onset.  
We assume homogeneity of the boundary temperatures, 
$T_{\rm top}$ and $T_{\rm bot}$, in the lateral $x$ 
direction.

In the experiments the aspect ratio was 57, 
so in the simulation \cite{Chiwata} 
the periodic  boundary  condition was imposed in the 
$x$ direction with period $4L$. 
 This  period was chosen  
because  the roll period is  
close to $2L$ slightly above the onset 
for infinite lateral dimension \cite{Chan}.   
Then,  in steady states in 
the region $1<Q/Q_{\rm on} \ls 5$,  the linear relation 
\be 
Q/ Q_{\rm on}-1 \cong A_0 
[ \Delta T /(\Delta T)_{\rm on}-1 ] 
\label{eq:3.1}
\en   
 was  numerically obtained  with $A_0 \cong  2.2$  
 in good agreement with  the experiments.  
From $Nu= [Q/\Delta T]/ [Q_{\rm on}/(\Delta T)_{\rm on}]$, 
the behavior of $Nu$ is known from  (3.2) 
in the range  $1<Q/Q_{\rm on} \ls 5$. In particular, 
slightly above the onset, we have 
\be 
Nu-1 \cong  A_1 ( Ra^{\rm corr}/Ra_{\rm c}-1) + \cdots  .
\label{eq:3.2}
\en 
where $A_1 \cong 0.64$ in 
fair agreement with the theoretical 
value ($A_1 \cong 0.70$ for $Pr=7.4$) \cite{Busse}.  
This behavior is also consistent with  (2.22).

In this work   we are interested in 
fluid motion  for relatively 
large $Ra$ up to $3 \times 10^6$. 
In the  following we show 
two sets of the numerical results. 
In the first set, periodic sidewalls 
are assumed  at $x=0$ and $x=L_\perp$ with period 
$L_{\perp}= 4L$ as in Ref.\cite{Chiwata}. 
In Table 1 the  steady state values of 
$\Delta T$, $Ra^{\rm corr}$, $Ra$,  $Nu$, 
and $\bar{R}e$  
are written,  where $\bar{R}e$  is a Reynolds number to be 
defined in (3.11).   They are obtained 
  for $Q=0.0458$ 
$\mu$W$/$cm$^2 (\cong  3.4 Q_{\rm on}$), 
$0.965$$\mu$W$/$cm$^2 (\cong  71 Q_{\rm on}$),  
and $122.2$$\mu$W$/$cm$^2 
(\cong  9 \times 10^3  Q_{\rm on}$). 
For the smallest  $Q$  the system 
tends to a time-independent convective state,  
as  already studied in 
 Ref.\cite{Chiwata}, while for the other   values  of $Q$ 
the system tends to a chaotic state without macroscopic 
boundary shear flow. 
In the second set, 
we  perform simulations 
for  $A=1,2,$ and 3 with  insulating and rigid 
sidewalls at $x=0$ and $AL$, at which $\bi v=0$ 
and through which there is no heat flux 
($\p \delta T/\p x=0$), as will be presented 
 in Figs.4, 12, and 13.

In addition, if the temperature difference 
will be  simply written as $\Delta T$, 
it  should be taken as 
the time average of $\Delta T(t)$ 
in a steady state. We also assume that 
 $Pr$ is considerably larger than 1 
 in the following   arguments.

\subsection{Transient behavior}

In Fig.1  we  show numerically calculated  
$\Delta T(t)= T_{\rm bot}(t) 
-T_{\rm top}$ 
for $Q=0.965$$\mu$W$/$cm$^2$   in (a) 
and for $Q=122.2$$\mu$W$/$cm$^2$ 
in (b). 
They  nearly coincide with the upper  broken curve 
without convection (${\bi v}={\bi 0}$) in the initial 
stage before the  maximum 
is attained.   The latter curve 
is calculated  from (2.8) as  
\be 
[\Delta T (t)]_0 = \frac{Q}{\lambda}\sqrt{\frac{Dt}{\pi}}
\bigg [4- \int_0^\infty \frac{ds}{\sqrt{\pi s}}\cdot
 \frac{1-e^{-s}}{s+t/t_1 } 
\bigg ], 
\label{eq:3.3}
\en 
where $t_1$ is defined  by (2.15) and 
the integral in the brackets behaves 
 as $(\pi t_1/t)^{1/2}$ 
for $t \gg t_1$ \cite{Onukibook}.
If the piston term is  absent  and ${\bi v}={\bi 0}$, 
(2.8) becomes the simple diffusion equation,  yielding   
$[\Delta T(t)]_0 = (2Q/\lambda)(Dt/\pi)^{1/2}$, 
which is about half of $[\Delta T (t)]_0$ in (3.3) 
for $t \gg t_1$ (see Fig.3 in Ref.\cite{Chiwata}). 
We also show the numerically calculated 
$\Delta T(t)$  at fixed pressure 
where the piston term is absent 
($\alpha_s=0$ in (2.8)) but ${\bi v}\neq {\bi 0}$. 
In (a) the experimental curve is 
 shown to have a lower peak and overdamp more slowly 
than in  our simulation. In (b) the selected value of $Q$ 
is in the region where  no overshoot 
was observed in the experiment.  
See also  Fig.11, where the numerical curves of $\Delta T(t)$ 
will be given for  other choices of the parameters.

In Fig.2  we show time evolution 
of the temperature profile at $Q=122.2$ $\mu$W$/$cm 
for periodic sidewalls. 
In (A) and (B) 
small-scale mushroom-like 
plumes are ejected from the bottom. 
In (C) and (D) they reach the top   and 
are flattened there. 
In this initial stage   
the typical raising speed $v_{\rm pl}$ is 
estimated as  $L/t_{\rm tr}$ 
where $t_{\rm tr}$ is the traversing time. 
From (A)-(C) we find that it   
is nearly equal to   the 
free-fall velocity $v_{\rm g}$ defined by 
\be 
v_{\rm g} = (Lg\alpha_p \Delta T)^{1/2}= (RaPr)^{1/2}D/L,  
\label{eq:3.4}
\en 
which is $2.37$ cm$/$s.
In this case the plumes leave the bottom at 
zero velocity and go upward with 
their velocity roughly of the form, 
\be 
{v_{\rm pl}}(t)= v_\infty \bigg 
[1-\exp[-(t-t_0)/t_{\rm vis}] \bigg ], 
\label{eq:3.5}
\en  
where $t_0$ is the departure time, 
 $t_{\rm vis} \sim \rho R^2/\eta$ is the viscous 
 relaxation time with  
 $R$ being  the plume size, and 
\be 
v_\infty \sim R^2g \rho\alpha_p\Delta T /\eta
\label{eq:3.6}
\en  
is the terminal velocity achieved by balance  
between the buoyancy and the viscous drag.   
 For $t_{\rm tr} \ll t_{\rm vis}$ 
 the viscous drag is negligible 
 and we have 
  ${v_{\rm pl}}(t) 
 \sim v_{\rm g}^2 (t-t_0)/L$  
 and  $t_{\rm tr} \sim L/v_{\rm g}$. 
Thus, if the initial velocity is much less than 
$v_{\rm g}$, the {\it free-fall} condition  becomes 
\be 
R/L  \gg (Pr/Ra)^{1/4}, 
\label{eq:3.7}
\en 
under which $v_\infty =
(R/L)^2(Ra/Pr)^{1/2}v_{\rm g} \gg v_{\rm g}$.   
In  Fig.2, 
 $R/L \sim 1/3$ and $(Pr/Ra)^{1/4} \sim 0.04$, 
so the above condition is satisfied.

With the arrival of the plumes the heat current 
increases at the top, because $T_{\rm top}$ is fixed, 
and a negative deviation 
of $\delta s$ is produced in a layer near  the top. 
As can be known from (2.13), the piston effect is then operative, 
resulting in a homogeneous lowering of the temperature in the whole 
cell. In the time region around (E) the fluid is 
vigorously mixed with high Reynolds numbers. 
More precisely,  the height-dependent Reynolds number 
$\hat{R}e (z,t)$ to be defined in (3.12) below  
is about $20$  except in the vicinity of  the boundaries.
A downward flow of  cooler fluid regions is then produced 
from the top. In the steady state (F), the temperature deviation 
 becomes considerably smaller than 
 in the transient states, 
 and  the localized 
boundary shear flows are produced 
between outgoing and incoming plumes 
with thickness $\ell_v$ much smaller  than $L$.

The overshoot is 
more clearly illustrated in Fig.3, 
which displays 
the average of ${\delta T} (x,z,t)$ 
taken in the $x$ direction, 
\be 
\overline{\delta T}(z,t) \equiv 
 \int_0^{L_{\perp}} \frac{dx}{L_{\perp}}
 \delta T(x,z,t), 
\label{eq:3.8}
\en  
for the points (A), (C), (E), and (F) in Fig.1b. 
As a characteristic feature, 
the temperature in the 
interior consists of  global changes   
due to the piston effect and bumps 
due to localized plumes.  
In (E)  the cooler  layer becomes thicker  temporarily 
near the top due to the excess heat flow.

In our simulation  the raising plumes 
leave the bottom and reach 
 the top nearly simultaneously, 
   resulting  in a homogeneous temperature change.  
(i) Not far above  the  onset 
 this  mechanism is the main cause of  the overshoot in 
 compressible fluids.  Note that a small peak  
  appears in $\Delta T(t)$ even in 
the fixed pressure case ($\alpha_s =1$) 
as shown in Fig.2 of Ref.\cite{Chiwata} 
and as  was  observed by 
Behringer and Ahlers \cite{BA}.  Furthermore, in Ref.\cite{Chiwata},  
 the time scale of the overshoot (from the maximum to the 
minimum of $\Delta T(t)$)  due to the piston effect 
was predicted to be   of order 
$t_D/(Ra^{\rm corr}/Ra_{\rm c}-1)$, where $t_D= L^2/4D$($\cong 50$ s) 
is  the diffusion time. This fairly agrees  with later analysis of 
the experimental data \cite{MeyerTr}.  
(ii) For much larger $Q$ such as 
those in Figs.1a and 1b, however, 
the downward flow from the top is also 
rapid enough to produce  large overshoot, 
as  demonstrated by the curves at fixed (height-dependent) 
pressure. Whether fixed is the volume or the pressure, 
the time scale of the overshoot  is of  the order of the 
traversing time  $L/v_{\rm g}$ of the plumes due to gravity.

As regards the overshoot behavior of $\Delta T(t)$,  
agreement between  
our simulation  and the experiment \cite{MeyerPRE} 
becomes worse with increasing $Q$. We point out the 
possibility that in the experiment  
a synchronous arrival of plumes at the top
might have not been  realized for very  large $Q$ 
or for very short $L/v_{\rm g}$ because of  large 
lateral dimensions of the cell used. That is, 
if some plumes arrive 
at the top and others leave the bottom at the same time, 
 negative interference between  currents up and 
down will suppress overshoot.

\subsection{Steady state behavior}

Now we discuss the Nusselt number 
$Nu$ in steady states. Fig.4 shows 
the combination 
$Ra^{\rm corr}(Nu-1)/(Ra^{\rm corr}-Ra_{\rm c})$ 
vs $Ra^{\rm corr}/Ra_{\rm c}-1$ for 
periodic sidewalls  and for $A=1,2,$ and 3.  
This combination depends on $Ra^{\rm corr}$ 
and $A$ from (2.34) in steady states. 
The data (solid line) \cite{MeyerPRE} excellently agree 
with the numerical results   
for  periodic sidewalls. 
 We find that the scaling relation 
(1.5) nicely holds for 
$Ra^{\rm corr}/Ra_{\rm c}\gs 10$   
for periodic sidewalls, 
while it holds only 
for $Ra^{\rm corr}/Ra_{\rm c}\gs 10^3$ at  
$A=1$.   The exponent $a$ in (1.5) is close to 
$2/7$,  but $a=1/4$ is also consistent 
with our numerical data. 
 If   $A\sim 1$  and $Ra^{\rm corr}$ is not 
very large such that the plume size is 
of order $L$, large-scale fluid motions are 
suppressed by the rigid  sidewalls.  
This marked tendency of 
the $A$ dependent crossover of $Nu$ 
was already reported in 
measurements for  $A=0.5, 1$, 
and $6.7$ \cite{Wu}. It was also confirmed in the   
3D simulation by Kerr \cite{Kerr1} for  periodic 
sidewalls.

In Fig.5 we show the  steady-state 
temperature deviation 
$\overline{\delta T}(z)$ averaged in the $x$ direction 
as in (3.8)  and in time 
for the three values of 
$Q$ in Table 1 for  periodic sidewalls 
with period $L_{\perp}= 4L$.  
The averages taken along the $x$ direction become  only 
weakly fluctuating in time in steady chaotic states 
(the relative fluctuations being of order $10\%$ 
for the largest $Q$). 
 As has been observed ubiquitously 
in the previous simulations, 
the temperature  gradient becomes localized within  
thermal boundary layers with thickness 
$\ell_T$.  
Because $\Delta T \cong  
2\ell_T Q/\lambda$ for $\ell_T \ll L$,  
it is related to $Nu$ by (1.6). 
The  arrows in Fig.5  
represent the maximum points, 
 $z=\ell_v$ and $L-\ell_v$, of the variance  of 
the horizontal  velocity   defined by 
\be 
{v}^*_x(z)= \bigg  
[  \int_0^{L_{\perp}}\frac{dx}{L_{\perp}}
  v_x(x,z,t)^2 \bigg ]^{1/2} . 
\label{eq:3.9}
\en  
In  Fig.6 we plot the normalized velocity 
variances,  ${v}^*_x(z)/v_{\rm g}$ in (a) and 
${v}^*_z(z)/v_{\rm g}$ in (b), where  
$v_{\rm g}$ is  defined by (3.4)    and 
\be 
{v}^*_z(z)= \bigg  
[\int_0^{L_{\perp}}
 \frac{dx}{L_{\perp}} v_z(x,z,t)^2 \bigg ]^{1/2} . 
\label{eq:3.10}  
\en 
The time average of $v_x^2$ and $v_z^2$ 
in the brackets is also taken in these 
figures.  On one hand,  ${v}^*_x$ take maxima 
at $z= \ell_v$ and $L-\ell_v$, where 
$\ell_v$ is hardly distinguishable from  
$\ell_T$. On the other hand, 
${v}^*_z$ is largest at the middle of the cell. 
We also find that the sum (the kinetic-energy variance)  
 $({v}^*_x)^2+ ({v}^*_z)^2$ is nearly  constant in the 
 interior,  which was a finding 
 reported in Ref.\cite{Kerr1}.  
At large $Ra$ the maxima of  ${v}^*_x$ 
and ${v}^*_z$  are  of the same order 
and will be identified as 
the typical  plume velocity $v_{\rm pl}$.
In our simulation we have 
$v_{\rm pl} \sim  
0.1 v_{\rm g} (\propto Ra^{1/2})$, 
which is consistent  
with velocity  measurements  
\cite{Sano,Tong}.

 Kerr and Herring \cite{Kerr2} made 
  similar plots of the height-dependent velocity variances 
 in their 3D simulations    for free-slip sidewalls. 
 They found  that   the characteristic 
length $\ell_v$ defined by the  peak positions  
of ${v}^*_x(z)$ becomes longer than 
$\ell_T = L/2Nu$ with increasing $Ra$;  
for example, for $Pr=7$ they obtained 
$\ell_v/\ell_T \sim 1$ at  $Ra=10^4$ 
and $\ell_v/\ell_T \sim 3$  at  $Ra=10^7$. 
Verzicco and Camussi   obtained 
a similar slow growing  of 
 $\ell_v/\ell_T$ at large $Ra$ for $Pr>1$ 
 in their 3D simulation with $A=1$ \cite{Ver}. 
Also similarly, our 2D simulation with  $Pr=7.4$ 
gives  $\ell_v/\ell_T = 2.54$ and $1.1$ for 
$Q=122.2$ and $0.965$ $\mu$W$/$cm, respectively, 
but we cannot draw 
a definite conclusion 
because of our limited range of $Ra$.

In Fig.7 we plot an overall Reynolds number $\overline Re$ vs 
$R^{\rm corr}/R_{\rm c}-1$ in the simulation 
for  periodic sidewalls. 	It is defined by 
\be 
{\overline Re}  =\frac{\rho}{\eta} 
 \bigg [ \av{|{\bi v}\cdot\nabla{\bi v}|^2} {\bigg  /}  
 \av{|\nabla^2 {\bi v}|^2} \bigg ]^{1/2}, 
\label{eq:3.11}  
\en   
where the averages are taken in the whole space region. 
The  ${\overline Re}$ is smaller than 1 
for $R^{\rm corr}/R_{\rm c}\ls 5$ \cite{Chiwata}. 
For larger values of  $R^{\rm corr}$, it 
exceeds 1 and the effective exponent 
$\p (\ln {\overline Re})/\p (\ln Ra^{\rm corr})$ 
is from $1/4$ to $1/3$. 
However, as suggested by Fig.6, 
the strength of the velocity fluctuations  
 strongly depends  on 
 the distance from the boundary, 
so  it is more informative to  
introduce a  height-dependent 
Reynolds number,  
\be 
\hat{R}e(z)=\frac{\rho}{\eta}
 \bigg [ \int_0^{L_{\perp}} d{x}
|{\bi v}\cdot\nabla{\bi v}|^2 \bigg / \int_0^{L_{\perp}} d{x}
|\nabla^2 {\bi v}|^2  \bigg ]^{1/2},  
\label{eq:3.12}  
\en  
where the time averages of  the integrands are taken.
As shown in Fig.8, $\hat{R}e(z)$ takes 
maxima at $z \sim \ell_v$ and $L-\ell_v$ of order 
\be 
\hat{R}e(\ell_v) \sim \ell_v v_{\rm pl} \rho/\eta, 
\label{eq:3.13}  
\en 
where $v_{\rm pl} \sim 
{v}^*_x(\ell_v)$.  This relation indicates 
$
\hat{R}e(\ell_v) \sim  Ra^{1/2-a}   
$ 
with $a \cong 2/7$ 
from $v_{\rm pl} \sim  0.1 v_{\rm g}$ 
and $\ell_v\sim \ell_T$.  
The $\hat{R}e(z)$ becomes considerably  smaller in the interior than 
at $z \sim \ell_v$, whose origin is the sparseness of 
the plumes in the interior (see (3.20) below).
We confirm that $\bar{Re}$ is of the order of the space 
average $\int_0^Ldz \hat{R}e(z)/L$.  In the literature 
\cite{SiggiaR,Siggia,Kadanoff,Ca,Gross,Gro}, however, 
the (large-scale) Reynolds number has been identified as 
$Re=v_{\rm pl}L \rho/\eta$, which is much larger 
than $\hat{R}e(\ell_v)$ in (3.13) by  $L/\ell_v$.  
(For roll patterns, as was discussed    below (2.21), 
we uniquely  have  $Re =  v_{\rm pl}L\rho/\eta$.)

At very large  $Ra$ 
the boundary layers  
should gradually crossover from a laminar state 
to  a turbulent state 
except within  thin viscous sublayers 
with thickness $z_{0}$ much shorter 
than $\ell_v$. In the inertial region 
$z_0 \ls z \ls \ell_v$ of the boundary layer, 
it is   natural to 
expect the logarithmic velocity profile  \cite{Landau_h}, 
\be 
{v}^*_x(z) = b_0^{-1}  (\sigma_0/\rho)^{1/2} 
 [ \ln (z/z_{0})  + c_0] , 
\label{eq:3.14}  
\en 
where  $\sigma_0$ is  the amplitude of the shear stress 
at the boundary with   $b_0$ 
and $c_0$ being dimensionless numbers  of order 1. 
We may  set  $\sigma_0 = 
\eta \lim_{z\rightarrow 0}D_{xz}(z)$, where  
$D_{xz}(z)$ is the variance of the velocity gradient, 
\be 
D_{xz}(z)= 
\bigg [ \int_0^{L_{\perp}} \frac{dx}{L_{\perp}}
 \bigg ( \frac{\p}{\p z} v_x(x,z,t)\bigg )^2 \bigg ]^{1/2} . 
\label{eq:3.15}
\en 
Then ${v}^*_x(z)\cong (\sigma_0/\eta)z$ 
as  $z \rightarrow 0$. It is appropriate to 
 define $z_0$ by \cite{Landau_h} 
\be 
z_0=  \eta/(\rho \sigma_0)^{1/2}, 
\label{eq:3.16}
\en
which ensures   $\hat{R}e(z_0) \sim 1$. 
The size of $\sigma_0$ should be equal to 
 the typical size of $\rho v_x v_z$ at $z=\ell_v$ 
 even if we consider localized  shear flows,  so 
 we also have  
 \be 
 \sigma_0 \sim \rho v_{\rm pl}^2.
\label{eq:3.17}
\en  
The ratio of the two lengths $z_0$ and $\ell_v$ is  
given by   
\be 
\ell_v/z_0 \sim v_{\rm pl} \ell_v
 \rho/\eta \sim \hat{R}e(\ell_v), 
\label{eq:3.18}  
\en 
which grows with increasing $Ra$. 
In  Fig.9a,  ${v}^*_x(z)$ is  fitted to 
the above logarithmic form in the inertial region for 
$Q=122.2$ $\mu$W$/$cm, where  
$(\sigma_0/\rho)^{1/2}
= 0.067 v_{\rm g}= 0.16$ cm$/$s, $b_0=1.2, c_0=0.9
7,$  
and $z_{0}= 0.025L$. In Fig.9b,  
we plot ${v}^*_x(z)$ and 
$zD_{xz}(z)$ 
on a logarithmic scale. 
We may conclude that these quantities do  
 not  behave as $z$ in the inertial region of the boundary 
 layers, although our $Ra$ is not large enough to 
unambiguously demonstrate the logarithmic velocity profile.  
Here we point out that our results are not consistent with 
Shiraiman and Siggia's primary assumptions of 
 $\ell_T<\ell_v$ 
and the linear profile of the mean shear 
flow,  $v_x \propto z$,  in the region 
$z<\ell_T$  \cite{SiggiaR,Siggia}.

In contrast to the averages 
taken along the $x$ direction, 
those taken along the $z$ direction 
are rapidly varying functions 
of  time  at large $Ra$ due to the random 
plume motions. We consider  the vertical velocity variance 
defined by 
\be 
{v}^*_z (x,t) 
= \bigg [\int_0^{L}\frac{dz}{L}
 {v}_z(x,z,t)^2 \bigg ]^{1/2}. 
\label{eq:3.19}
\en  
In Fig.10 we display  snapshots of ${v}^*_z (x,t)$, where  
the time average is not taken and peaks arising from the plumes  
become more apparent with increasing $Q$.  
For our $Ra$ realized,  the 
space regions occupied 
by the plumes become more   sparse 
with increasing $Ra$ in the interior. 
 As the plumes move through the cell, 
they remain  distinguishable 
from  the ambient fluid 
because the thermal 
diffusion length $(DL/v_{\rm pl})^{1/2}$ does not much 
exceeds $\ell_v$.  So we may 
define the  volume fraction of the plumes  
$\phi_{\rm pl}$.     The convective heat current 
is of order  
$\phi_{\rm pl}v_{\rm pl} C_p\Delta T \sim \lambda Nu \Delta T/L$, 
leading to 
\be 
\phi_{\rm pl} \sim  D/\ell_T  v_{\rm pl},   
\label{eq:3.20}
\en 
which is of order  $Pr^{-1} \hat{R}e(\ell_v)^{-1} \ll 1$ 
from (3.13). For much larger $Ra$, 
the plumes will generate 
smaller scale eddies,  ultimately 
leading to fully developed turbulence 
in the interior, as will be discussed in Section 4.

\subsection{Overall Temperature Fluctuations}

When  a    plume 
 with a volume $V_{\rm fl}$ reaches  the  boundary, 
 it transfers a heat  of order $C_p \Delta T 
V_{\rm fl}$ to the  boundary wall. As indicated by 
(2.13),  the piston effect 
then gives rise to a homogeneous 
change in  $\av{\delta T}(t)$ of order  
\be 
(\delta T)_{\rm fl} \sim \gamma_s  (V_{\rm fl}/V) \Delta T .  
\label{eq:3.21}
\en  
Of course,  the real plumes are extended objects 
and are continuously arriving at the  
boundary in high $Ra$ convection.  
Thus $V_{\rm pl}/V$ in the above formula 
should be regarded as the fluctuation amplitude 
of the plume volume fraction $\phi_{\rm pl}$ 
 in the interior, although we do not know 
 its dependence on  $Ra$ etc at present. 
If $T_{\rm top}$ and $Q$ at the bottom are fixed 
as  in our simulation, 
$\Delta T(t)$ should 
also consist of  fluctuations of the same origin. 
Because of the strong critical divergence  of $\gamma_s$,  
we expect that the relative amplitude 
$(\delta T)_{\rm fl}/ \Delta T$ would increase
as $\epsilon$ is decreased with a fixed size of $\Delta T$.

Fig.11 displays   
time sequences of  $\av{\delta T}(t)$ and $\Delta T(t)$ 
at fixed volume and pressure 
for periodic sidewalls with $L_\perp=4L$, 
which demonstrates 
 strong correlations 
between these two deviations at fixed volume.   
 In case  (a) (upper figure) we set  
 $\epsilon=0.05$ ($\gamma_s =22.8)$, 
$\Delta T= 0.17 $mK, $Pr=7.4$,  $Ra=
7.38 \times 10^4$, and $Nu=4.06$, 
while in case (b) (lower figure) 
we set $\epsilon=0.01$ ($\gamma_s =119)$, 
$\Delta T= 0.19 $mK, $Pr=37.7$, 
$Ra=4.14 \times 10^5$, and $Nu=6.04$.  
 The steady state values of $\Delta T$ in the two cases 
 are chosen to be  only slightly different. 
At fixed volume,  the fluctuations  of 
$\av{\delta T}(t)$ and $\Delta T(t)$ 
are strongly correlated,  
and are  larger and slower for (b)  than for (a) 
in steady states $(t \gs 100)$.  This is because
of  the critical enhancement of the piston effect and the 
critical decrease of $D$ with decreasing $\epsilon$. 
At fixed pressure, 
where the piston effect is absent, 
  $\Delta T(t)$  exhibits noises 
  much smaller than those at fixed volume and 
    $\av{\delta T}(t)$ smoothly changes in time.      
It is worth noting that  
this noise increase at fixed volume accompanied with 
an increase of  $Ra$   is contrary 
to the usually measured noise behavior 
of the temperature.  For  non-critical fluids,   if 
the  temperature is measured at the center  of a  cell, 
its  fluctuation amplitude 
 divided by $\Delta T$   
is known to decrease  with increasing $Ra$  as 
$Ra^{-\beta_n}$. The exponent $\beta_n$ 
was about  $0.15$ in a cell with $A=1$ 
\cite{Kadanoff,N0}.

\subsection{Random Reversal of Macroscopic Flow}

For  a convection cell with $A \sim 1$,   
it is well-known that 
large-scale shear flow develops 
near the boundary of the cell 
for large enough $Ra$ \cite{N1,Cioni,Tong}.  
Moreover, it has also been observed that 
 the global circulation changes its orientation 
over long time scales \cite{Cioni,N1}. 
For the case of $A=1$, $\epsilon=0.05$, 
$Q=40.7$ $\mu$W$/$cm$^2$, $Ra=1.68\times 10^6 (\cong Ra^{\rm corr})$, 
and $Nu =5.97$,   
 we plot  a numerical time sequence of a 
circulation $\Gamma (t)$ in Fig.12.   Here,  
\bea 
\Gamma (t) &=&
\int_d^{L-d} dx [ v_x(x,L-d,t)-v_x(x,d,t)]/L \\
&+&\int_d^{L-d} dz [ v_z(L-d,z,t)-v_z(d,z,t) ]/L,   
\label{eq:3.22}
\ena  
where the integration is along a square contour  
with distance $d=0.05L$ from the cell boundary. 
This quantity is 
positive for clockwise circulation 
 and negative for counterclockwise circulation.  
 In Fig.12,   $\Delta T(t)$ is also plotted, 
 which  exhibits particularly large fluctuations 
 on the occasion of orientation changes. 
 This is a natural result because large-scale 
  reorganization of the flow pattern 
 is needed for an orientation change.   
 Fig.13 illustrates 
   the velocity patterns  at 
$t=228, 269$, and $311$s in Fig.12. 
 They  closely resemble   
a picture of the measured velocity pattern 
 in Ref.\cite{Tong}.

\setcounter{equation}{0}
\section{Scaling theory}

Rayleigh numbers realized in 
the existing  simulations are still moderate in the sense that  
 the plumes do not  have enough kinetic energies such 
  that they do not generate fully developed 
turbulence in the interior. 
In this {\it pre-asymptotic}  regime of steady states, 
we may understand the numerical 
and experimental data using 
a  very simple {\it zeroth-order} theory.  
First, we set $\ell=\ell_T = \ell_v$ 
neglecting  the possible small 
difference between $\ell_T$ and $\ell_v$ 
mentioned below (3.10). The plume sizes in the horizontal 
direction are also of order $\ell$. 
Second, in our simulation 
the plumes are ejected into the interior 
with a velocity $v_{\rm pl}$,  
for which the viscous drag and the 
buoyancy are balanced or  
\be 
\eta \ell^{-2}v_{\rm pl} 
\sim g\alpha_p\Delta T.
\label{eq:4.1}
\en 
Thus $v_{\rm pl}$ is of the order of the terminal velocity 
$v_\infty \sim Ra D\ell^2/L^3$ in (3.6) with $R \sim \ell$. 
In the interior we find that   (i) 
gravity-induced  acceleration of the plumes  
is suppressed by  the viscous drag,   
(ii) $(v^*_x)^2+(v^*_z)^2$ 
is nearly independent of $z$
  as stated below (3.10), and 
(iii) the last two terms 
on the left  hand side of (2.17) 
are numerically of the same order.  For example, the ratio of  
the average of  $(\alpha_pg\delta T)^2$ 
in the $x$ direction to  that of 
 $ |(\eta/\rho)\nabla^2{\bi v}|^2$ 
is about 4 at $z \sim \ell$ and is fluctuating 
around 1 in the interior  for the largest $Q$ in 
Table 1.  These support $v_{\rm pl} 
\sim v_\infty$ in the interior.
Third, to the sum rule (2.36) for the velocity gradients, 
the  contribution from the 
boundary layers is  
 of order $v_{\rm pl}^2/\ell L$, 
while that from the  interior 
is of order   $\phi_{\rm pl} 
v_{\rm pl}^2/\ell^2 \sim D v_{\rm pl}/\ell^3$ 
from (3.20).  If use is made of (4.1) and the sum rule 
(2.36),   these  boundary-layer  
and bulk  contributions 
become  both of order $RaNu(D/L^2)^2$,   
which  has  also been confirmed numerically. Thus, 
\be 
v_{\rm pl} \sim Ra^{1/2}D/L,   
\label{eq:4.2}
\en 
\be  
Nu \sim L/\ell \sim 1/\phi_{\rm pl}\sim Ra^{1/4}. 
\label{eq:4.3}
\en 
These quantities are independent of $Pr$. 
In particular, the independence of $Nu$ on $Pr$ is consistent with 
the experiments \cite{Cha,Ahlers-Xu,MeyerPRE}.
Note that  $v_{\rm pl} (\sim v_\infty) $ is smaller than 
$v_{\rm g}$ in (3.4) by  $Pr^{-1/2}$.

Our height-dependent Reynolds number 
at $z = \ell$ in (3.13) becomes 
\be 
\hat{R}e (\ell) \sim Ra^{1/4}/Pr . 
\label{eq:4.4}
\en 
The usual large-scale Reynolds number is given by 
$Re \sim v_{\rm pl}L\rho/\eta \sim Ra^{1/2}/Pr$. 
As  $\hat{R}e (\ell)$ exceeds a crossover value  $Re^*$, 
plumes will induce turbulence in the interior. 
Our  simple scaling theory 
is valid for $Ra \ls (Re^*Pr)^4$. 
In our simulation we have  $\hat{R}e (\ell) \cong 
0.38 Ra^{1/4}$, so that 
if we set $Re^* \sim 10^3$ (regarding  plumes as 
jets \cite{Landau_h}), 
the upper bound is crudely estimated as  $5\times 10^{13}$. 
The transition from 
the scaling (4.3) to the asymptotic  
scaling occurs over a very wide  
range of $Ra$. Similarly, Grossmann and Lohse \cite{Gross} 
considered a transition of a laminar boundary-layer flow 
to a turbulent boundary layer 
when the local Reynolds number on the scale of $\ell_v$ 
at $z \sim \ell_v$ 
exceeds a value of order $420$. 
Then  $Nu$ was claimed to be better 
expressed by 
\be 
Nu \sim 
Ra^{1/4}(1+ C_1 Ra^{b}) 
\en 
than the single power-law form, 
where  $C_1$ and $b$ are  small coefficient 
and exponent, respectively,   
dependent on $Ra$ and $Pr$ under investigation 
(both being of order $0.1$). 
This  proposed form of $Nu$ was later 
claimed to be in good agreement   with 
data  \cite{Ahlers}.

Here it would be informative to add 
more supplementary explanations of  
the previous scaling theories. 
(i) Shiraiman and 
Siggia \cite{SiggiaR,Siggia} 
assumed fully developed turbulence 
in the interior. Then the maximum of 
the turbulent velocity gradient is 
  of order $S_{\rm d}= \eta k_d^2/\rho= 
  (v_{\rm pl}^3\rho/L\eta)^{1/2}$ 
 at the smallest eddy size $k_d^{-1}$ 
 ($ \sim (\eta/\rho v_{\rm pl})^{3/4}L^{1/4})$ 
  if the Kolmogorov 
cascade is assumed with the energy 
dissipation rate $v_{\rm pl}^3/L$ \cite{Landau_h} 
(the sparseness of ejected plumes 
being neglected).  
If the left hand side of 
the sum rule (2.36)  is estimated as $S_d^2$, 
it follows the relation,  
\be 
v_{\rm pl} \sim (Pr Nu Ra)^{1/3}D/L . 
\label{eq:4.6}
\en 
Furthermore, they  assumed the linear    
horizontal velocity profile 
$v_x \sim (\sigma_0/\eta)z$ in the region 
$z\ls \ell_T \sim L/Nu$, where $\sigma_0$ 
is given by (3.17). 
From the thermal diffusion equation 
$v_x \p \delta T/\p x = 
D\nabla^2\delta T$ (the time-dependent 
fluctuations being neglected), they obtained  
the scaling, 
\be 
\ell_T^{-3} \sim \rho v_{\rm pl}^2/\eta D L , 
\label{eq:4.7}
\en 
by setting $\p/\p x \sim L^{-1}$ 
and $\nabla^2 \sim (\p/\p z)^2 \sim \ell_T^{-2}$ 
for a cell with $A \sim 1$.  From 
(4.6) and (4.7) they found 
\be 
Nu \sim Pr^{-1/7} Ra^{2/7}. 
\label{eq:4.8}
\en 
However, as discussed below  (3.14), 
our simulation suggests that 
the velocity deviates significantly 
from the linear profile in  the 
boundary layers. 
(ii) Castaing {\it et al}. \cite{Ca} assumed the 
balance (4.1) at the length  $\ell_T$, 
\be 
v_{\rm pl}\sim \ell_T^{2}g\alpha_p\Delta T/\eta 
\sim RaNu^{-2}D/L.
\label{eq:4.9}
\en  
They furthermore  assumed that  the typical 
temperature scale in the interior is 
 $(\delta T)_c \sim v_{\rm pl}^2/\alpha_p gL$ 
and that the average 
heat current ($\cong Nu \lambda \Delta T/L$)  
 is of order  $C_p(\delta T)_c v_{\rm pl}$. From these relations 
 $(\delta T)_c$ may be eliminated to give (4.6).  
If we combine (4.6) and 
(4.9), we are again led to  (4.8). 
Therefore,  to justify their arguments, the presence of 
fully  developed  turbulence in the interior  seems to be 
required.  (iii) Grossmann and Lohse \cite{Gross,Gro}
estimated the bulk and boundary-layer 
contributions to the sum rules for the temperature gradient and 
the velocity gradients, 
the incompressible version of (2.35) and 
(2.36). Their primary  assumption is 
that the boundary layer thickness for 
the velocity is given by $\ell_v  \sim L/Re^{1/2}$ in terms of 
the large-scale Reynolds number $Re$ \cite{Landau_h}.   Note that this 
assumption is not consistent with our zeroth-order scaling theory 
with respect to the $Pr$ dependence. 
In particular, in the case 
where $Pr>1$  and  the boundary-layer contributions 
are dominant both for the temperature and the velocity, 
they obtained  $Nu \sim Pr^{-1/12}Ra^{1/4}$. 
In this case we also find 
 $\ell_v/\ell_T \sim Pr^{1/3}$ from their theory. 
 They predicted that 
this pre-turbulent  scaling   
 crossovers to the asymptotic turbulent scaling  very slowly  
as in (4.5).

\section{Concluding Remarks}

We have presented a hydrodynamic model of compressible fluids 
properly taking  into account the piston effect and the 
 adiabatic temperature gradient effect. 
  Though performed in two dimensions, 
our simulation has revealed some new effects 
unique in near-critical fluids, 
 such as the overshoot 
behavior and the amplification  of the overall temperature 
fluctuations as $T\rightarrow T_c$.   
It generally explains  
the experimental findings \cite{MeyerPRL,MeyerPRE}, 
but  a discrepancy remains in 
 the overshoot behavior at high heat flux $Q$ 
 as discussed in Section 3. 
It is desirable to extend simulation to  
smaller $\epsilon$ and  higher $Ra$.  
Also more  experiments  on  the  
overshoot and the temperature 
noises etc. are needed  to resolve the discrepancy 
and to confirm  the new predictions.  
As by-products, we have numerically examined 
steady state properties not treated in the previous 
simulations,  such as the logarithmic 
velocity profile and the random reversal 
of macroscopic shear flow.  They  are 
universal aspects present both in compressible and 
 incompressible fluids.

We have assumed that the fluid is  in the supercritical 
region not very close 
to the critical point such that 
the conditions (2.1) and (2.2) are satisfied. 
However, if $\Delta T$ 
exceeds $T-T_{\rm c}$ or if $T_{\rm top}$ 
is below $T_{\rm c}$,  we encounter   
a variety of new effects  
such as boiling and wetting under heat flow and gravity 
\cite{Onukibook,Onukiboil}.  We believe that 
such problems should  provide us a new challenging field 
in which nonlinear dynamics and phase transition 
dynamics are coupled.  
These problems  are beyond the scope of this paper.

We  thank H.  Meyer 
for valuable suggestions and  comments. 
Thanks are also due to P. Tong for informative 
correspondence. 
This work is supported by Japan Space Forum 
 grant H13-264.

\end{multicols}

\vspace{3mm}

\begin{table}
\caption{Parameters at $\epsilon=0.05$ 
in steady states for periodic sidewalls}
 \label{tab:simulation}
\begin{tabular}{cccccc}
$Q$ ($\mu$W$/$cm s) & $\Delta T$ (mK)  & $Ra^{\rm corr}$ & $Ra$ 
 & $Nu-1$  & ${\overline Re}$  \\
\hline
0.0458 & 0.0154 & 3.43$\times 10^3$ & 6.69$\times 10^3$ & $0.714$ & $0.655$ \\
\hline
0.965  & $0.135$ & $ 5.87\times 10^4 $ & $ 5.54\times 10^4 $ & $3.04$ & $3.035$ \\
\hline
122.2 & $6.89$  & 2.91$\times 10^6$ & 2.91$\times 10^6$ & $9.29$ & $7.89$ \\
\end{tabular}
\end{table}

 Fig.1. 
$\Delta T (t)$ vs time  (solid line) calculated from (2.8) 
and (2.17)   for (a) $Q=0.965$ $\mu$W$/$cm$^2$ 
and  (b) $Q=122.2$ $\mu$W$/$cm$^2$. 
The temperature profiles
for the points ($\Box$) on the curve in (b)  
 are given in Fig.2. 
The experimental data ($+$) \cite{MeyerPRE}   are shown in (b).  
The upper broken curves in (a) and (b)  represent 
the theoretical result  (3.3)  obtained from 
integration of (2.8) with ${\bi v}={\bi 0}$. 
The dotted curves  represent 
the numerical ones  in the fixed pressure condition 
without the piston effect. 
\\

Fig.2.  
Temperature profiles at (A), (B), (C), (D), (E), and (F)     
on the curve of  $Q=122.2$ $\mu$W$/$cm$^2$ in Fig.1b ($\Box$). 
The temperature (and  velocity) deviations 
are more enhanced in the transient states 
(A)$\sim$(E) than in a steady state (F). 
The $\delta T$ at the bottom boundary $z=0$ 
is equal to $\Delta T (t)$ in Fig.1b. 
The plumes tend to be  connected between  bottom 
and top because  $Pr=7.4$. 
\\

Fig.3. 
Time evolution of $\overline{\delta T}(z,t)$ 
defined by (3.5)  at the points (A), (C), (E), and (F) in Fig.1b 
for  $Q=122.2$ $\mu$W$/$cm$^2$.   
\\

 Fig.4.
 Numerical results of 
 $Ra (Nu-1)/(Ra^{\rm corr}-Ra_{\rm c})$ 
 vs  $Ra^{\rm corr}/Ra_{\rm c}-1$ 
 in steady states,  obtained under 
 the periodic boundary condition ($+$) and  
 for $A=3 (\Box),  2 (*),$ and $1(\times)$. 
 The   first curve ($+$) 
 is close to the experimental results 
 for $A=57$ \cite{MeyerPRE}  (solid line)
 and is well fitted to the 
 scaling form (1.5) with $a \cong 2/7$ for 
  $Ra^{\rm corr}/Ra_{\rm c} \gs 10$ .  
 With decreasing the  aspect ratio $A$,   
 crossover to the scaling 
 occurs at much larger  $Ra^{\rm corr}$.  
\\

Fig.5. 
Height-dependent average temperature 
profiles  $\overline{\delta T}(z)$ 
divided by $\Delta T$ in steady states for the 
three $Q$ values in Table 1. 
The  arrows represent 
the maxima of ${v}^*_x(z)$ in Fig.6a.\\

Fig.6.  Normalized height-dependent 
variances, ${v}^*_x(z)/v_{\rm g}$ for the horizontal velocity  in (a) 
and ${v}^*_z(z)/v_{\rm g}$ 
for the vertical velocity in (b) in steady states   
 for the three $Q$ values in Table 1.  \\

Fig.7. Oveall Reynolds number $\overline Re$ 
defined by (3.11) as a function of 
 $Ra^{\rm corr}/Ra_{\rm c}-1$  
 in steady  states for  $Q=122.2$ $\mu$W$/$cm$^2$.
\\

Fig.8. Height-dependent 
Reynolds number $\hat{R}e(z)$ 
defined by (3.9) in steady states
 for   the three $Q$ values in Table 1.\\

Fig.9. (a) Height-dependent  
velocity variance ${v}^*_x(z)$ 
defined by (3.9)   on a semi-logarithmic scale 
in steady states
  for  $Q=122.2$ $\mu$W$/$cm$^2$. 
 (b)  ${v}^*_x(z)$  (upperline) and 
 velocity gradient variance $zD_{xz}(z)$  
 defined by (3.15) (lower line) on a logarithmic scale. 
\\

Fig.10. Snapshots of 
the normalized velocity variance ${v}^*_z (x,t)/v_{\rm g}$ 
averaged in the $z$ direction defined by (3.17) for the three 
values of $Q$ in Table 1. The system is periodic with 
period $4L$ in the $x$ direction. 
The peak heights increase with increasing $Q$.
For the largest  $Q$ 
this quantity changes in time  
as the plumes move in the cell, 
while for the other $Q$ it  
is weakly dependent on or independent 
of time.  
\\

Fig.11.  $\av{\delta T}(t)$  and 
$\Delta T(t)$ at fixed volume (solid line) 
and at fixed pressure (broken line) 
 for $\epsilon=0.05$ (upper figure) and $0.01$ (lower figure). 
The noises of these quantities 
at fixed volume increase as the reduced temperature 
$\epsilon$ is decreased. 
\\

Fig.12. 
Time evolution of the circulation $\Gamma (t)$ 
defined by (3.22) 
(upper figure) and $\Delta T(t)$ (lower figure) for 
$Q=122.2$ $\mu$W$/$cm$^2$ in a cell 
with  $A=1$. The orientation of the 
macroscopic flow changes  on a  
time scale of 50 s.  The sign of $\Gamma (t)$ 
represents the orientation of the macroscopic 
circulation, while the fluctuations of 
$\Delta T(t)$ become large when the orientation changes. 
 \\

Fig.13.  
Velocity patterns  at 
$t=228, 269$, and $311$ s  for the run  in Fig.12.  
At $t=228$ s the orientation is counterclockwise,  
while at $t=311$ s  it  is clockwise. 
At $t=269$ s  two large 
eddies with different 
orientations can be seen.\\


\begin{references}
\bibitem{one} The usual hydrodynamic equations 
for one-component fluids in 
the Boussinesq approximation 
  are valid only when (i) the specific heat ratio 
$C_p/C_V$  is close to 1, (ii) $\Delta T$ is much larger than  
$a_{\rm g} L$ in (1.2), and (iii) the gravity-induced 
stratification is weak such that (2.2) holds.
\bibitem{gra} For $0<\epsilon \ll 1$ 
the scaling form 
$\alpha_p = \epsilon^{-\gamma} G(u)$ holds 
with  $u=  (\rho/\rho_{\rm c}-1)/\epsilon^{\beta}$. 
For  $|u| \ll 1$ or under (2.2) in gravity, 
we have  $G(u) \cong G(0)$ and $\alpha_p \cong G(0) 
\epsilon^{-\gamma}$.
\bibitem{homo} To be precise, the time constant of 
the pressure-homogenization is given by 
$t_{\rm hom}= (t_{\rm ac}t_1)^{1/2}$ under the condition 
$t_{\rm ac}\ll t_1$ in terms of the acoustic time 
$t_{\rm ac}=L/c$ and the piston time $t_1$ \cite{Onukibook}. 
Here the damping of the pressure oscillation is 
mainly caused by the damping of the oscillatory 
heat current in the thermal diffusion  layers.
\end{references}
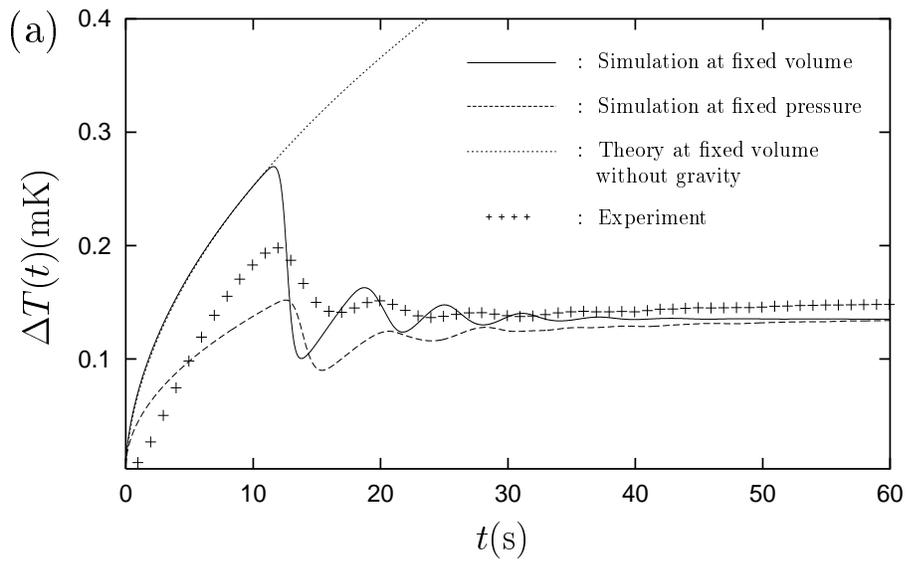
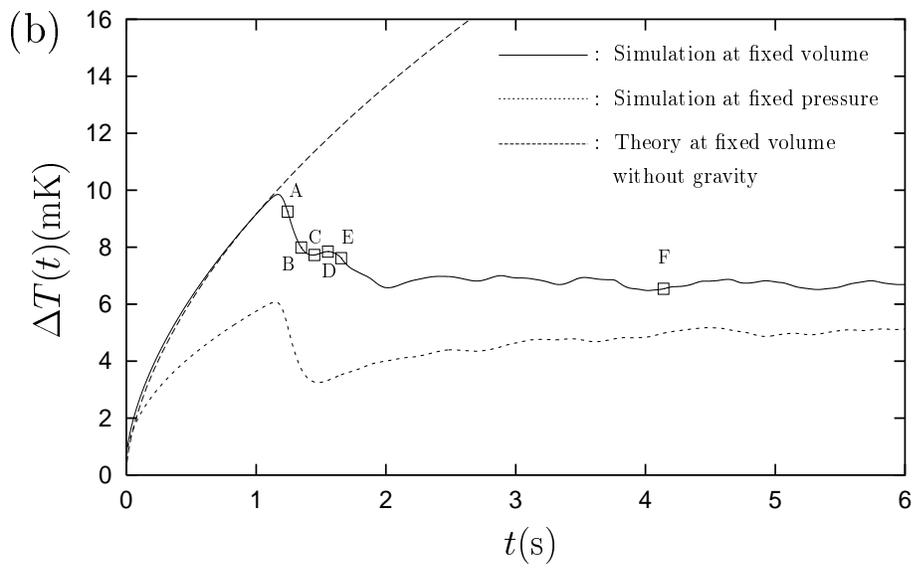
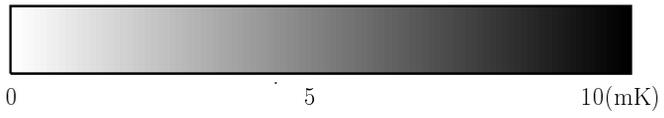
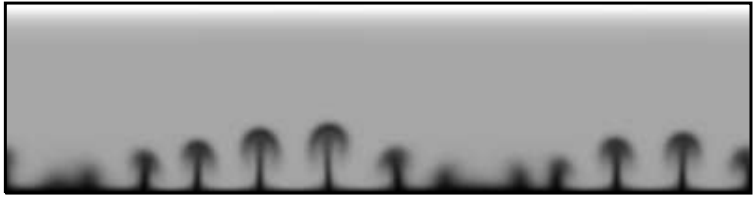
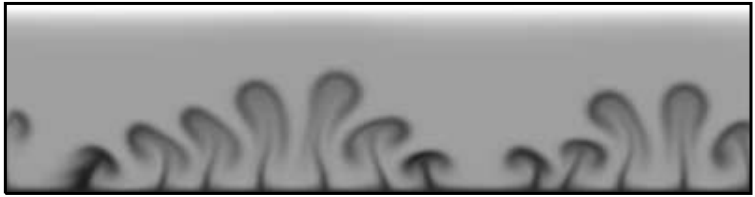
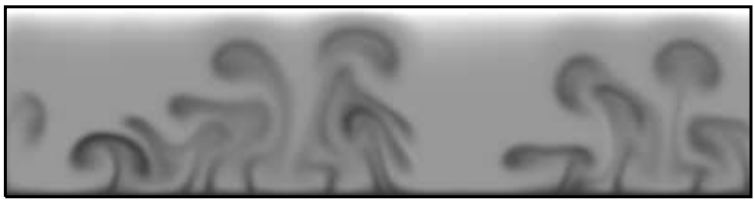
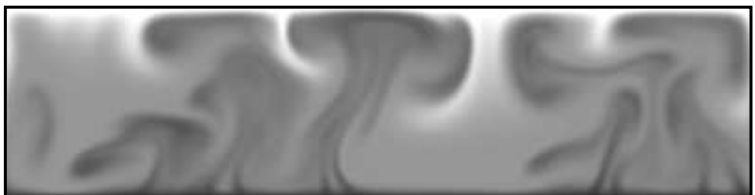
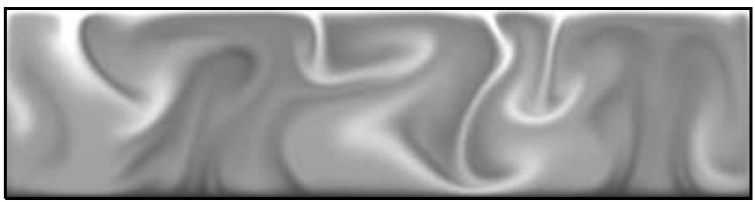
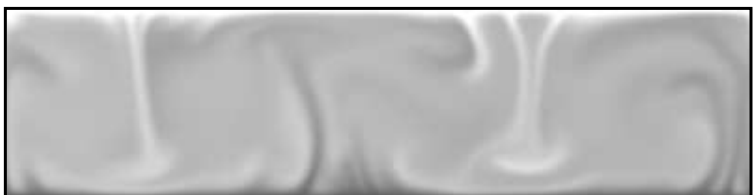
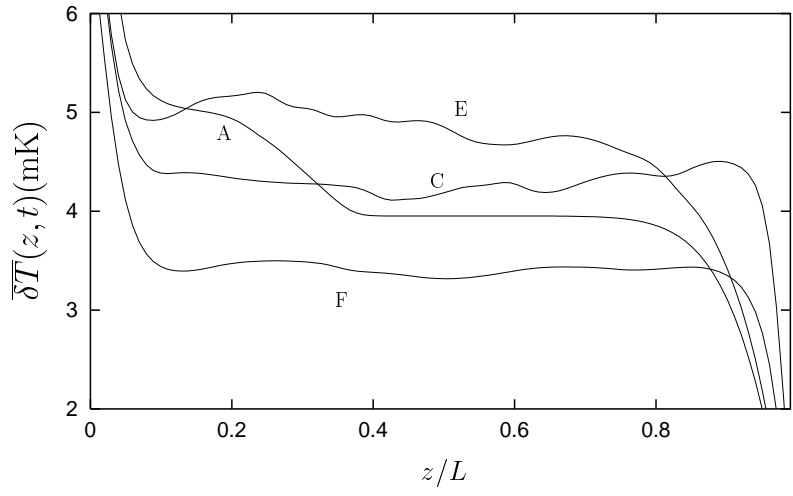
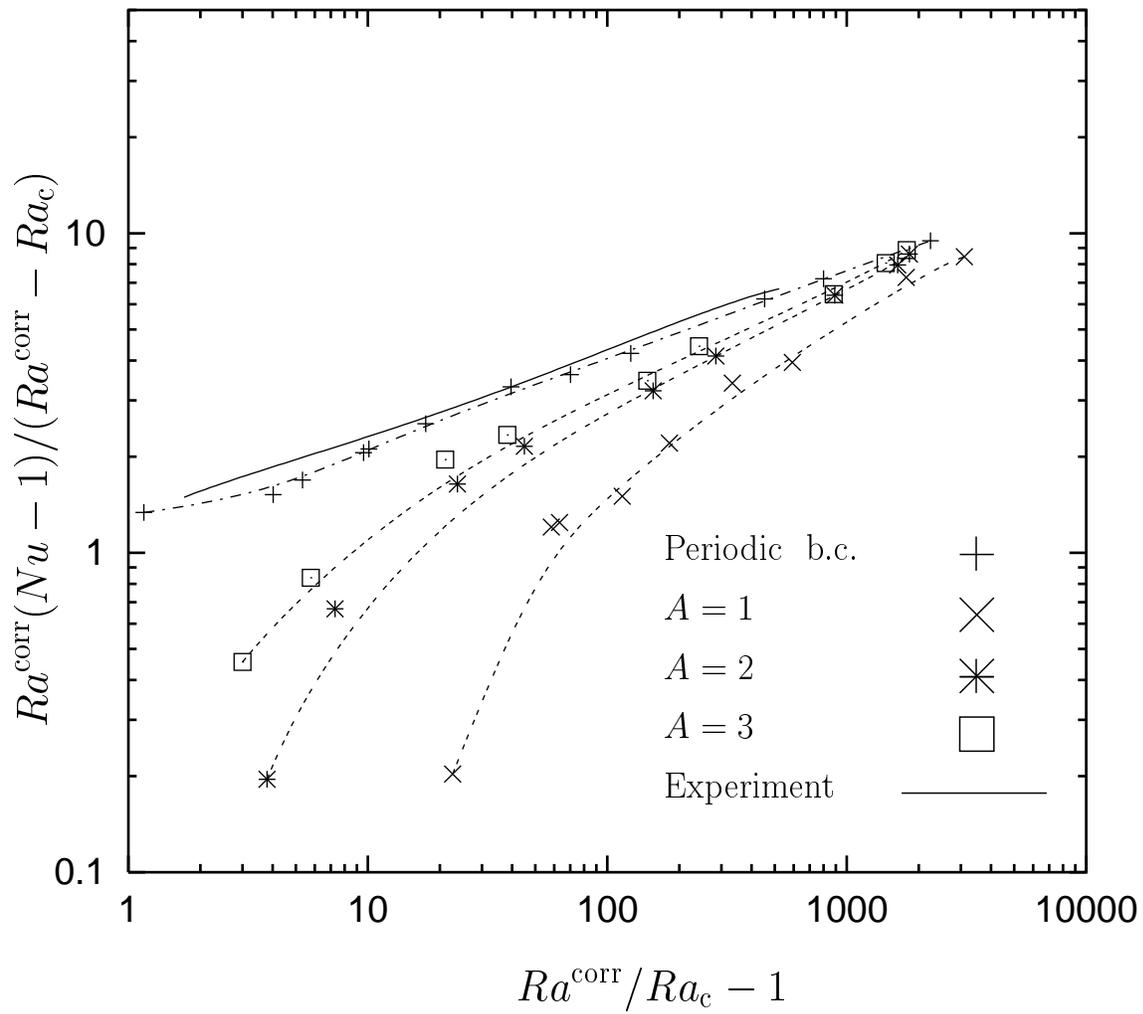
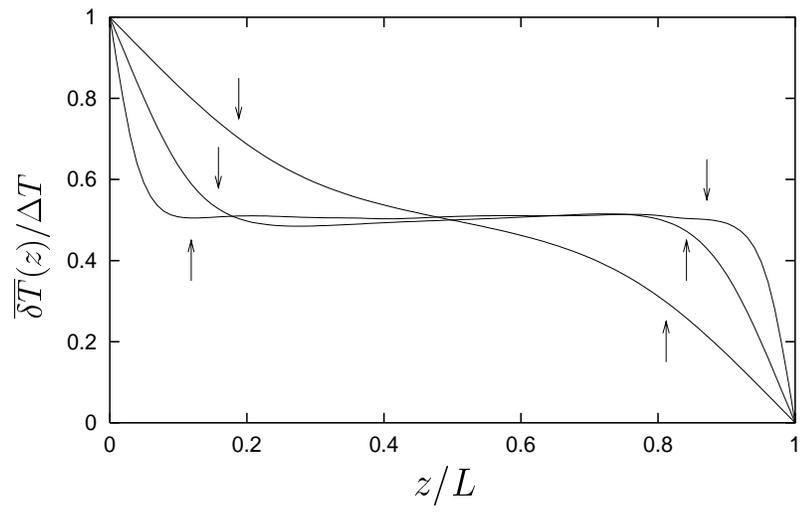
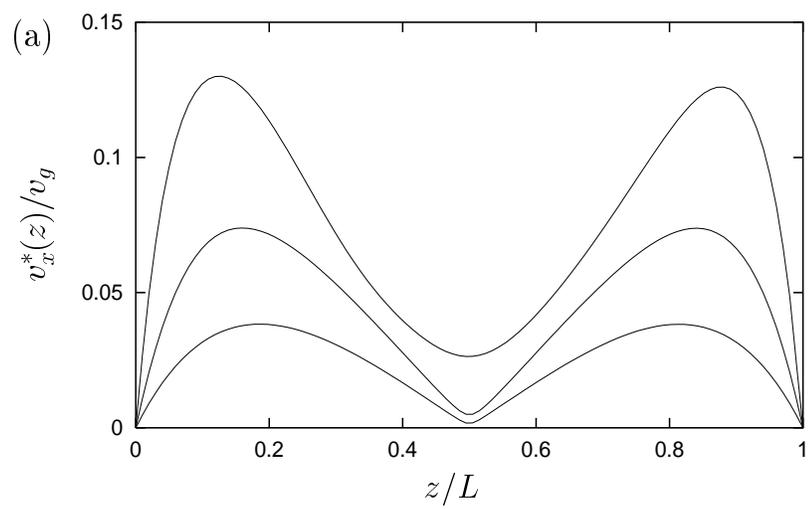
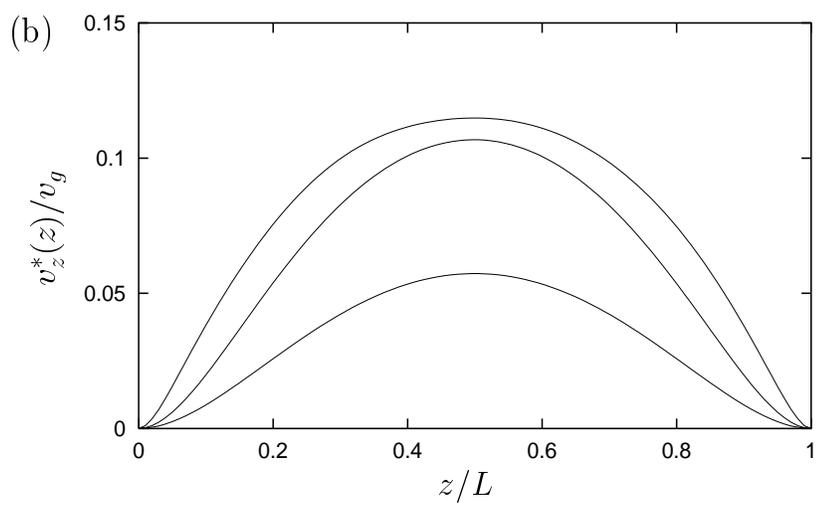
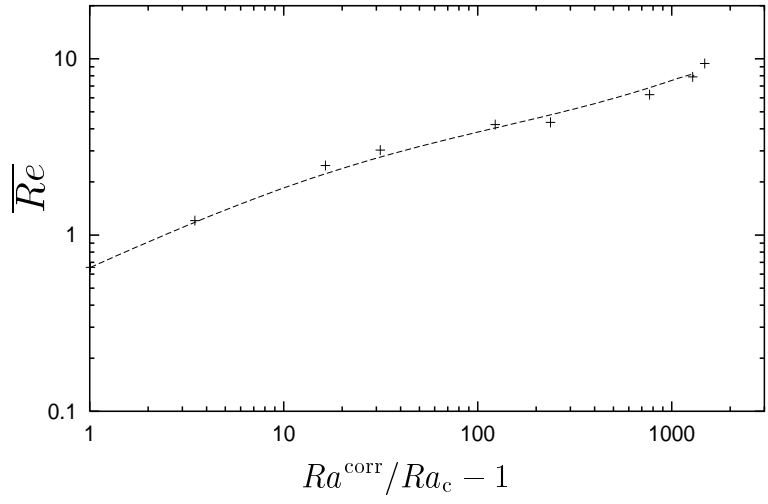
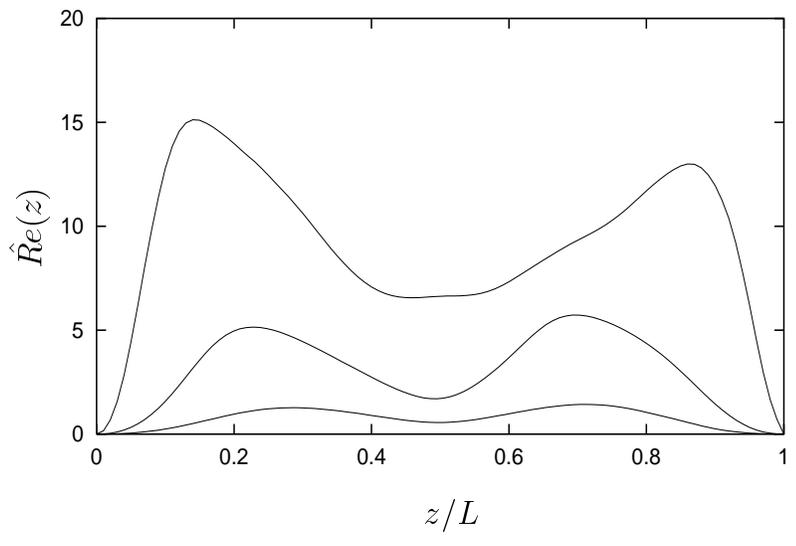
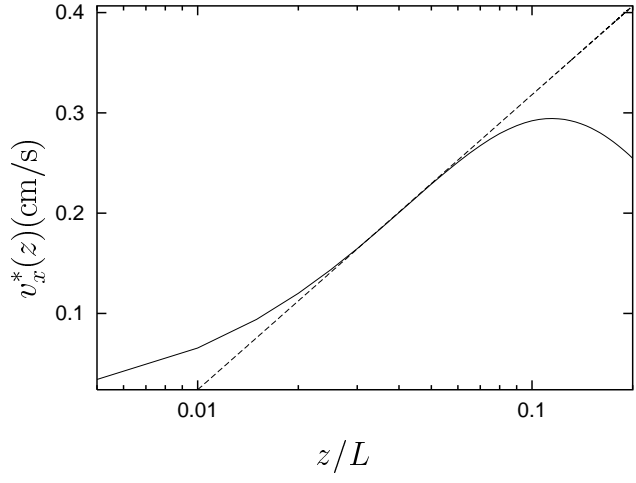
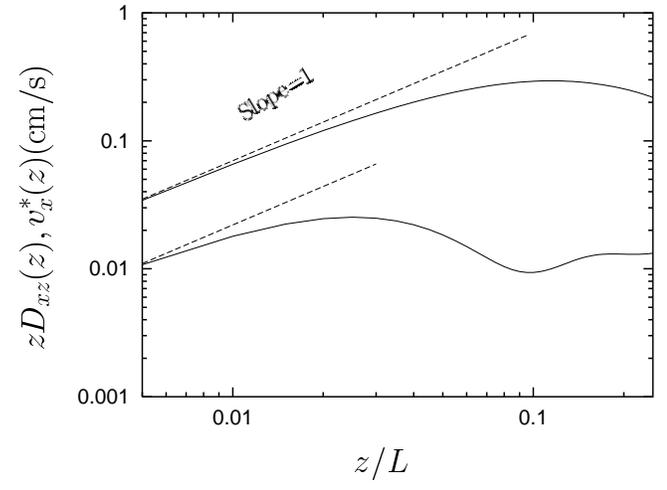
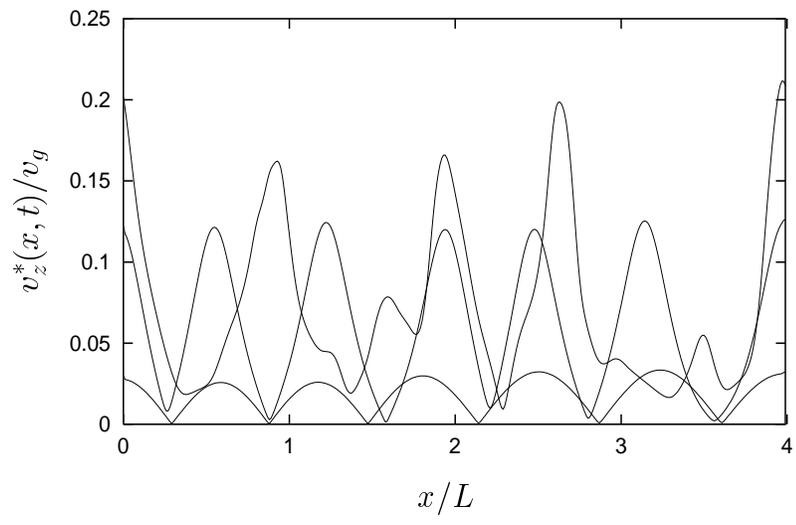
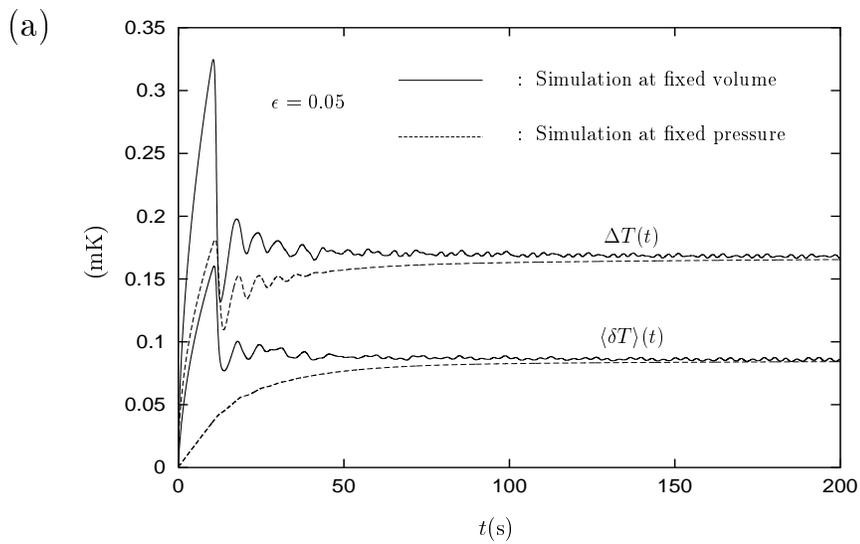
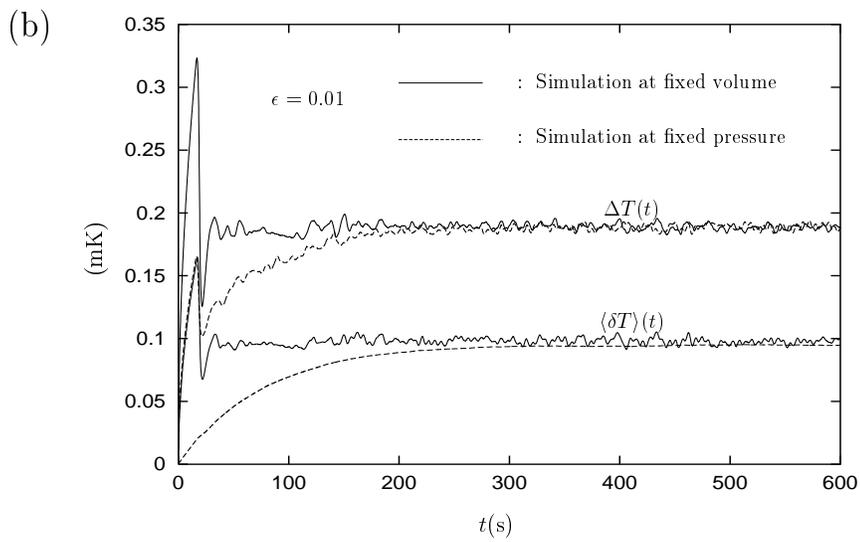
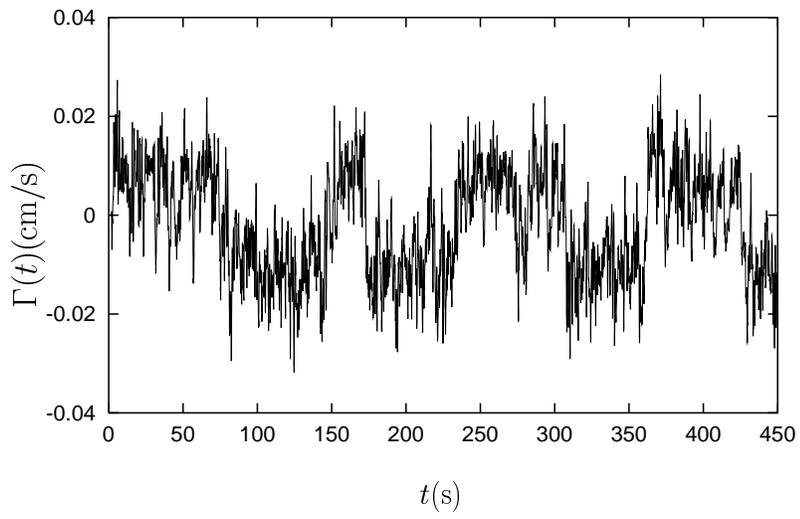
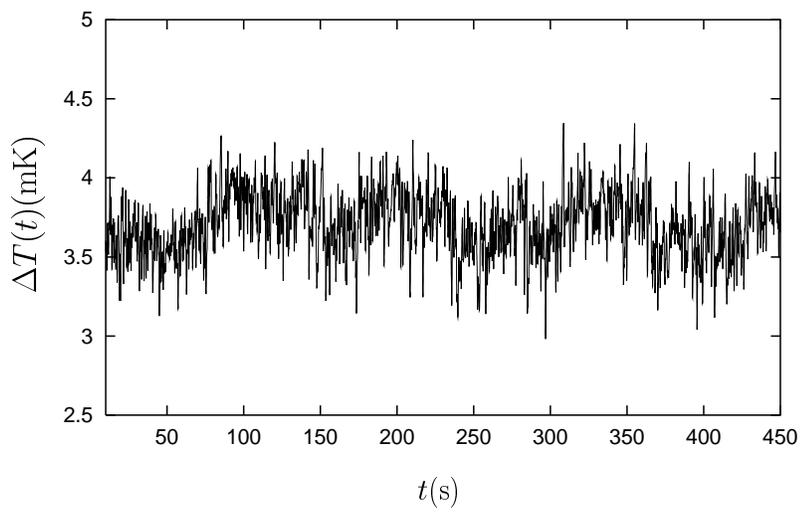
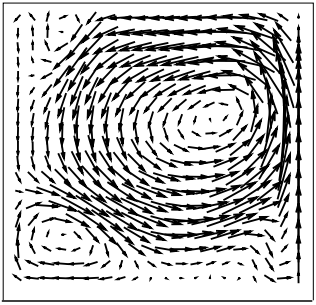
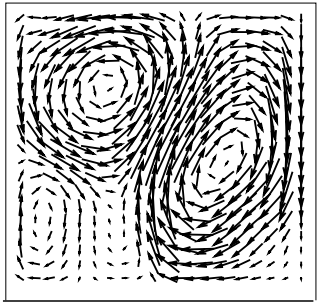
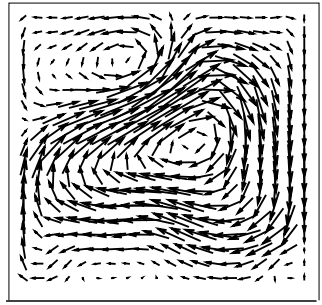
\end{document}